\documentclass[twocolumn,groupedaddress,assymb,amsmath]{revtex4}
\usepackage{}
\usepackage{graphicx}
\usepackage{amssymb}
\usepackage{amsmath}

\usepackage{extarrows}
\usepackage[bookmarks=false]{hyperref}
\hypersetup{colorlinks=true,citecolor=blue,linkcolor=blue,urlcolor=blue,pdfstartview=FitH,bookmarksopen=true}

\begin{document}
\title{Shortcuts to adiabaticity with general two-level non-Hermitian systems}
\author{T. Z. Luan,$^{1}$ H. Z. Shen,$^{1,2,}$\footnote{Corresponding author: shenhz458@nenu.edu.cn} and X. X. Yi$^{1,2}$}
\affiliation{$^1$Center for Quantum Sciences and School of Physics,
Northeast Normal University, Changchun 130024, China\\
$^2$Center for Advanced Optoelectronic Functional  Materials
Research, and Key Laboratory for UV Light-Emitting Materials and
Technology of Ministry of Education, Northeast Normal  University,
Changchun 130024, China}
\date{\today}

\begin{abstract}
Shortcuts to adiabaticity (STA) are alternative fast processes which reproduce the same final state as the adiabatic process in a finite or even shorter time, which have been extended from Hermitian systems to non-Hermitian systems in recent years, but they are barely explored for general non-Hermitian systems where off-diagonal elements of the Hamiltonian are not Hermitian into account. In this paper, we propose a shortcuts to adiabaticity technique which is based on transitionless quantum driving algorithm to realize population transfer for general two-level non-Hermitian systems and give both exact and approximate analytical solutions of the corresponding counteradiabatic driving Hamiltonian, where the latter can be extended to the zeroth-order and first-order terms by applying perturbative theory. We find that the first-order correction term is different from the previous results, which is caused by the non-hermiticity of the off-diagonal elements. We work out an exact expression for the control function and present examples consisting of general two-level system with gain and loss to show the theory. The results suggest that the high-fidelity population transfer can be implemented in general non-Hermitian systems by our method, which works even with strong non-hermiticity and without rotating wave approximation (RWA). Furthermore, we show that the general Hamiltonian whose off-diagonal elements are not conjugate to each other can be implemented in many phyisical systems with the present experimental technology, such as atom-light interaction system and whispering-gallery microcavity, which might have potential applications in quantum information processing.
\end{abstract}

%³É

\maketitle
\section{Introduction}
Manipulating the state of a quantum system with time-dependent interacting fields is a fundamental operation in atomic and molecular physics with applications such as laser-controlled chemical reactions, metrology, interferometry, nuclear magnetic resonance or quantum information processing \cite{Sawicki4217,Vitanov52763,Kral7953,Law761055,Kuhn69373,Zheng080502}. Among the powerful and interesting strategies \cite{Torosov233001,Brif075008}, the quantum adiabatic theorem \cite{Born51165,Kato5435} offers a simple way to prepare and manipulate quantum states in principle in a robust way, which is ubiquitous in many physics systems \cite{Zhu097902,Muga175501,Demirplak1168028,delCampo031606,delCampo100502,Tian153604}. In quantum mechanics, an ``adiabatic process" is a slow change of Hamiltonian parameters that keeps the populations of the instantaneous eigenstates constant \cite{Bergmann701003}. These processes are frequently chosen to drive or prepare states in a robust and controllable way and have also been proposed to solve complicated computational problems. The main drawback is that they are slow by definition. As the system remains in the instantaneous eigenstates, there is no heating or friction, but the long operation times needed may render the operation useless or even impossible to implement because decoherence would spoil the intended dynamics. Therefore, it is natural to look for novel methods which are robust and fast to improve or take instead of the adiabatic methods.

A compromise is to use speeded-up shortcuts to adiabaticity \cite{Chen123003,Torrontegui62117,GueryOdelin045001}, which may be broadly defined as the processes that lead to the same final populations as the adiabatic approach but in a shorter time. There are different motivations for the speedup that depend on the setting. In optics, time is often substituted by length to quantify the rate of change so the shortcuts imply shorter, more compact optical devices. In mechanical engineering, we look for fast and safe protocols, say of robotic cranes, to enhance productivity. In microscopic quantum systems, slowness often implies decoherence, the accumulation of errors and perturbations, or even the escape of the system from its confinement. The shortcuts to adiabaticity provide a useful toolbox to avoid or mitigate these problems and thus to develop quantum technologies. There are different approaches to engineer shortcut to adiabaticity, including transitionless quantum driving (also known as counteradiabatic driving) \cite{Demirplak1079937,Demirplak1096838,Demirplak154111,Berry365303,Ibanez062136,Chen052109,Takahashi062117,Opatrny015025,Saberi060301,Chen023405,Ibanez100403,
Song023001,Song052324,Zhang7384,Deffner021013,Campo115703,Chen38484,Wang075201,Kang022304,Ge025207,Li063411,Vacanti053017}, ``fast-forward" scaling \cite{Torrontegui013601,Masuda043434,Masuda062108,Masuda4661135,Zhu023307}, inverse engineering based on Lewis-Riesenfeld (LR) invariants \cite{Lewis101458,Chen062116,Chen063002,Chen033405,Lohe42035307,Lu105201,Luo062127,Torrontegui043408,Gungordu062312,Kang5291700004,Lu22262,Bartolotta06752,Gao3626,Kiely115501,
Laforgue023415,Chen023841} and so on \cite{Ibanez043402,MartinezGaraot043402,Chen062319,Zhong682016}. The robustness and near perfect fidelity nature of the adiabatic processes are preserved in these shortcuts to adiabaticity techniques. As these methods provide arbitrarily fast dynamics, they are less vulnerable to the decoherences, decays and effects of interaction with the environment. Over the past decade, these methods have been explored rigorously across various branches of physics such as waveguide couplers \cite{Lahini193901,Sun34280,Tseng2121224,Paul053406,Ke50393}, Bose-Einstein condensates \cite{Schaff9323001,Drummond063619,DupontNivet053420,Ozcakmakli055001,Muga241001,Campo60005,Masuda013626}, entangled state preparation \cite{Marr033817,Paul052303,Xia012326,Zhang015202,Yang012326,Wang70162,Wu633669,Wu746255,Xu090301,Chen012325,Ji032323,Kang052311,Lu045206}, quantum transport \cite{GueryOdelin063425,Zhang043410,An12999,Corgier055002,Torrontegui013415,Chen043415,Chen013631,Ding063410}, state preparation for quantum information process \cite{Santos012311,Opatrny023815}, wavepacket splitting \cite{Torrontegui033630}, many-body state engineering \cite{MartinezGaraot213001}, non-Hermitian systems \cite{Torosov052502,Torosov063412,Li7114,Song2421674,Li2530135,Ibanez023415,Chen5301700247,Mostafavi050404} and so on \cite{Ruschhaupt093040,Mukherjee062108,Torrontegui033605,Chen053403,Lu023627,Lu063414,Chen023624,Yu062317,Zhu052203,Huang053313,Ding040401,MunueraJavaloy054054,Ban206602,Xu023125,
Sala043623,Deffner012001,Song022332,Baksic021801,Longhi062122,Schaff113017,Chen033856,Huang012333,Chen032328,Kang042336,Wang062118,Wu043413,Liang2323798,Takahashi150602,Funo150603}. It is worth to note that an increasing interest has been devoted to study non-Hermitian Hamiltonians because a non-Hermitian Hamiltonian, such as, a Hamiltonian obeying PT-symmetric \cite{Bender5243,Bender70957,Mostafazadeh43205,Mostafazadeh367081}, could produce a faster than Hermitian evolution while keeping the eigenenergy difference fixed \cite{Bender040403,Uzdin415304}. For non-Hermitian systems, the usual approximations and criteria are not necessarily valid \cite{Wu053421}, therefore the arguments and results that are applicable for Hermitian systems have to be reconsidered and modified \cite{Ibanez033403}.

In the past few years, shortcuts to adiabaticity methods have been generalized to non-Hermitian systems and show us a possibility to speed up quantum population transfer without changing coherent control fields. However, most works devoted to two-level non-Hermitian system shortcuts to adiabaticity studied the case in which the off-diagonal elements of the Hamiltonian are conjugate to each other \cite{Li2530135,Song2421674,Ibanez023415}. These motivate us to explore feasible shortcuts to adiabaticity in general two-level non-Hermitian systems, where off-diagonal elements of the Hamiltonian are not Hermitian into account.

Moreover, it is worth mentioning that many interaction systems can be described by non-Hermitian Hamiltonians whose off-diagonal elements are not conjugate to each other \cite{Chen548192,Wiersig063828,Wiersig012119,Wiersig203901,Peng123842,Wang111610,Lai57665,Holler1809}. Recently, the experimental realization of anti-PT-symmetric optics by introducing a novel coupling mechanism has been reported \cite{Peng123842}. The time-dependent Hamiltonian in Ref.~\cite{Peng123842} is a general non-Hermitian Hamiltonian and its off-diagonal elements are not conjugate to each other. In addition, the general non-Hermitian Hamiltonian can also be obtained in two-state model with fixed azimuthal mode number in whispering-gallery microcavity \cite{Chen548192,Wiersig063828,Wiersig012119,Wiersig203901}. Therefore, it is natural to look for novel methods which are robust and fast to improve or take instead of the adiabatic methods in general two-level non-Hermitian systems.

In this paper, based on the above considerations, we propose a shortcuts to adiabaticity technique for general non-Hermitian two-level systems whose off-diagonal elements are not conjugate to each other. By using transitionless quantum driving method, we analytically derive both exact and approximate counteradiabatic driving Hamiltonians which steer the dynamics along the instantaneous eigenstates of the reference Hamiltonian. For the sake of clearness, we apply this method to a Gaussian model as an example to verify the validity of our general non-Hermitian shortcuts to adiabaticity and determine the exact control to speed up the adiabatic population transfer. We numerically compare the population transfers implemented in weak and strong non-hermiticity regimes under the control of the exact and approximate total Hamiltonians which consist of the reference Hamiltonian and the corresponding exact or approximate counteradiabatic driving Hamiltonian, respectively. We also apply this method to the two-level system with gain and loss and without RWA. We numerically calculate the population transfer dynamics with and without counter-rotating terms, respectively. Furthermore, the possible physical implementations of the general non-Hermitian two-level Hamiltonians with off-diagonal elements which are not conjugate to each other are also discussed.

The remainder of the paper is arranged as follows. In Sec. II, we study the shortcuts to adiabaticity based on transitionless driving algorithm applied to the general non-Hermitian two-level system and derive the exact and approximate counteradiabatic driving Hamiltonians, respectively. In Sec. III, we present the model to describe the system under study and discuss the population transfer controlled by total Hamiltonians with exact and approximate counteradiabatic driving Hamiltonians respectively in both weak and strong non-hermiticity regimes. The population transfer dynamics with RWA and without RWA are also discussed in this section. In Sec. IV, we discuss the experimental feasibility of the selected model and analyze the specific examples. Finally, we conclude with a summary of the paper in Sec. V.

\section{Basic theories}
In this section, we present a shortcuts to adiabaticity technique for an open two-level quantum system via transitionless driving algorithm, where the non-Hermitian Hamiltonian whose off-diagonal elements are not conjugate to each other typically describes subsystems of a larger system \cite{Muga395357}, which can be generally described by the Hamiltonian
\begin{eqnarray}
\begin{aligned}
{H_0}\left( t \right) = \hbar \left( {\begin{array}{*{20}{c}}
{h\left( t \right)}&{\Omega \left( t \right)}\\
{g\left( t \right)}&{e\left( t \right)}
\end{array}} \right),
\label{general_H0}
\end{aligned}
\end{eqnarray}
where $h\left( t \right)$, $\Omega \left( t \right)$, $g\left( t \right)$ and $e\left( t \right)$ are general time-dependent complex functions. The Hamiltonian (\ref{general_H0}) describes the general two-level non-Hermitian system, which can denote the specific physical systems under specific parameters, such as a atom-light interaction system whose coupling between two spatially separated probe fields is mediated through coherent mixing of spin waves \cite{Peng123842}, a two-state model with fixed azimuthal mode number in whispering-gallery microcavity \cite{Chen548192,Wiersig063828,Wiersig012119,Wiersig203901}, non-Hermitian Su-Schrieffer-Heeger (SSH) model \cite{Jiang052116,Yin052115}, a two-level system interacting with a circularly polarized light \cite{Joshi042117,Slavcheva053804}, and non-Hermitian systems with gain and loss \cite{Torosov052502,Torosov063412,Li7114}, etc.

The instantaneous eigenvalues for the Hamiltonians ${H_0}(t)$ and $H_0^\dag (t)$ given by (\ref{general_H0}) are respectively ${E_n}(t)$ and $E_n^ * (t)$, in which the corresponding instantaneous eigenstates respectively read $| {{E_n}(t)}\rangle $ and $| {{{\tilde E}_n}(t)} \rangle $, where the details of the derivations can be found in Appendix~\ref{APPA}. Before studying the shortcuts to adiabaticity, we need to make the adiabatic approximation, where the adiabaticity condition for time-dependent non-Hermitian Hamiltonian (\ref{general_H0}) is given by Eq.~(\ref{conditionform}) (see Appendix~\ref{APPB} for the details). Based on this, we can expand the evolution state $\left| {\Psi (t)} \right\rangle $ determined by Eq.~(\ref{SchH0}) in terms of the instantaneous eigenstate $\left| {{E_n}(t)} \right\rangle $ of Hamiltonian (\ref{general_H0}) as follows
\begin{equation}
\begin{aligned}
\left| {\Psi (t)} \right\rangle  = \sum\limits_n {{\alpha _n}} (t){e^{ - \frac{i}{\hbar}\int_0^t {{E_n}({t_1})d{t_1}} }}\left| {{E_n}(t)} \right\rangle
\label{expand_solution},
\end{aligned}
\end{equation}
where the coefficient ${{\alpha _n}} (t)$ satisfies
\begin{small}
\begin{equation}
\begin{aligned}
{{\dot \alpha }_n}(t) =&  - {\alpha _n}(t)\langle {{\tilde E}_n}(t)|{{\dot E}_n}(t)\rangle  - \sum\limits_{m \ne n} {{\alpha _m}(t)} \langle {{\tilde E}_n}(t)|{{\dot E}_m}(t)\rangle &\\
&\times \exp [ - \frac{i}{\hbar }\int_0^t {{G_{mn}}(t')dt']},&
\label{solution_step}
\end{aligned}
\end{equation}
\end{small}
with ${G_{mn}}(t) = {E_m}(t) - {E_n}(t)$. From the adiabatic theorem, we can know that there are no transitions in the adiabatic evolution of the system, that is to say, the evolution of ${\alpha _n}(t)$ only depends on ${\alpha _n}(t)$ (the first term on the right-hand side of Eq.~(\ref{solution_step})) while not on other probability amplitudes, which means the contribution of the second term on the right-hand side of Eq.~(\ref{solution_step}) approximates to zero. Integrating Eq.~(\ref{solution_step}) gets
\begin{equation}
\begin{aligned}
{\alpha _n}(t) = {e^{i{\gamma _n}(t)}}{\alpha _n}(0)
\label{solution},
\end{aligned}
\end{equation}
where ${\gamma _n}(t) = i\int_0^t {\langle {{{\tilde E}_n}(t')} |}{{{\dot E}_n}(t')} \rangle dt'$ denotes the geometric phase, which is generally complex function (depending on the instantaneous eigenstates) that generalize the real Berry phase of Hermitian systems and only relate to the path of the quantum state in Hilbert space. In addition, from the above analysis, the adiabaticity condition (\ref{general_condition}) can be obtained. We take the initial state as $\left| {\Psi (0)} \right\rangle  = \left| {{E_m}(0)} \right\rangle $, together with Eq.~(\ref{expand_solution}), which leads to ${\alpha _n}(0) = {\delta _{nm}}$. Therefore, with Eq.~(\ref{solution}) and the adiabaticity condition (\ref{general_condition}), we can obtain the adiabatic approximate wave function
\begin{equation}
\begin{aligned}
\left| {\Psi (t)} \right\rangle  = {e^{i{\gamma _m}(t)}}{e^{ - \frac{i}{\hbar}\int_0^t {{E_m}(t_1)d{t_1}} }}\left| {{E_m}(t)} \right\rangle
\label{solution_final},
\end{aligned}
\end{equation}
which is the solution to the Schr${\rm{\ddot o}}$dinger equation (\ref{SchH0}) with non-Hermitian Hamiltonian ${H_0}(t)$ in Eq.~(\ref{general_H0}).

Now our aim is to find a new Hamiltonian $H(t)$ whose solution of Schr${\rm{\ddot o}}$dinger equation should be consistent with that of adiabatic approximation. In the following, we shall explore the situation that the adiabaticity condition (\ref{conditionform}) fails and then apply the transitionless quantum driving algorithm to remedy the problem and achieve rapid evolution.

In the transitionless quantum driving algorithm, a Hamiltonian $H\left( t \right)$ is designed so that the adiabatic approximation for the time-dependent wave function evolving with a reference Hamiltonian ${H_0}\left( t \right)$ given by Eq.~(\ref{general_H0}) becomes exact. In order to realize the shortcuts to adiabaticity, we first write out two orthogonal projection operators corresponding to the relevant instantaneous eigenstates:
\begin{equation}
\begin{aligned}
{M_ + }\left( t \right) &= \left| {{E _ + }\left( t \right)} \right\rangle \langle {{{\tilde  E }_ + }\left( t \right)} |&\\
 &= \frac{1}{{S_ + ^2\left( t \right)}}\left( {\begin{array}{*{20}{c}}
{{{\Omega \left( t \right)g\left( t \right)}}}&{\Omega \left( t \right){\Lambda _ + }\left( t \right)}\\
{{\Lambda _ + }\left( t \right){g\left( t \right)}}&{\Lambda _ + ^2\left( t \right)}
\end{array}} \right),\\
{M_ - }\left( t \right) &= \left| {{E _ - }\left( t \right)} \right\rangle \langle {{{\tilde  E }_ - }\left( t \right)} |&\\
 &= \frac{1}{{S_ - ^2\left( t \right)}}\left( {\begin{array}{*{20}{c}}
{{{\Omega \left( t \right)g\left( t \right)}}}&{\Omega \left( t \right){\Lambda _ - }\left( t \right)}\\
{{\Lambda _ - }\left( t \right){g\left( t \right)}}&{\Lambda _ - ^2\left( t \right)}
\end{array}} \right),
\label{M}
\end{aligned}
\end{equation}
where ${\Lambda _ \pm }\left( t \right)$ and ${S_ \pm }\left( t \right)$ are given by Eq.~(\ref{Edgzz}). With Eqs.~(\ref{M}) and (\ref{Edg}), the counteradiabatic driving Hamiltonian which is required in the transitionless quantum driving algorithm \cite{Berry365303} can be constructed as
\begin{small}
\begin{equation}
\begin{aligned}
{H_1}\left( t \right) &= i\hbar \left[ {\frac{{{M_ + }\left( t \right){\partial _t}{H_0}{M_ - }\left( t \right)}}{{{E_ - }\left( t \right) - {E_ + }\left( t \right)}} + \frac{{{M_ - }\left( t \right){\partial _t}{H_0}{M_ + }\left( t \right)}}{{{E_ + }\left( t \right) - {E_ - }\left( t \right)}}} \right]&\\&
 = \frac{{i\hbar }}{{4{C_1}\left( t \right)}}\left( {\begin{array}{*{20}{c}}
{ - {B_1}\left( t \right)}&{A\left( t \right)}\\
{ - D\left( t \right)}&{{B_1}\left( t \right)}
\end{array}} \right),&
\label{H1_general}
\end{aligned}
\end{equation}
\end{small}where $A\left( t \right) = \dot \Omega \left( t \right)[e\left( t \right) - h\left( t \right)] - \Omega \left( t \right)[\dot e\left( t \right) - \dot h\left( t \right)]$, $D\left( t \right) = \dot g\left( t \right)[e\left( t \right) - h\left( t \right)] - g\left( t \right)[\dot e\left( t \right) - \dot h\left( t \right)]$, ${B_1}\left( t \right) = \Omega \left( t \right)\dot g\left( t \right) - \dot \Omega \left( t \right)g\left( t \right)$, and ${C_1}\left( t \right) = {d^2}\left( t \right) + \Omega \left( t \right)g\left( t \right)$.

Based on Eqs.~(\ref{general_H0}) and (\ref{H1_general}), we can obtain Schr${\rm{\ddot o}}$dinger equation $i\hbar | {\dot \Psi (t)} \rangle  = {H}\left( t \right)\left| {\Psi (t)} \right\rangle$ with $H\left( t \right) = {H_0}\left( t \right) + {H_1}\left( t \right)$ and $\left| {\Psi (t)} \right\rangle$ given by Eq.~(\ref{solution_final}), which drives the system along the adiabatic paths defined by ${H_0}\left( t \right)$ but beyond the adiabatic limit. That is to say, the transitionless quantum driving algorithm proposes procedure by adding an exact additional counteradiabatic driving Hamiltonian ${H_1}\left( t \right)$ given by Eq.~(\ref{H1_general}) to the initial general two-level non-Hermitian Hamiltonian ${H_0}\left( t \right)$ given by Eq.~(\ref{general_H0}), which drives the initial state $\left| {{E_m}(0)} \right\rangle $ to the final state $\left| {\Psi (t)} \right\rangle$ given by Eq.~(\ref{solution_final}) without final excitation. This nullifies the effect of non-adiabatic terms to drive the system exactly along the adiabatic path and is faster than the reference adiabatic process. Apparently, this method requires full knowledge of the instantaneous spectral properties (i.e., instantaneous eigenstates and eigenvalues) of the original system to compute the counteradiabatic driving Hamiltonian. For the atomic two-level systems, ${H_1}\left( t \right)$ will involve auxiliary laser or microwave interactions. The optional addition of ${H_0}\left( t \right)$ will imply different physical implementations.

The result of ${H_1}\left( t \right)$ in Eq.~(\ref{H1_general}) is different from the precious one \cite{Ibanez062136,Chen023405,Ibanez023415}, where ${B_1}\left( t \right)$ and ${C_1}\left( t \right)$ are both general complex functions. Also, $A\left( t \right)$ is neither complex conjugate to $D\left( t \right)$ nor equal to $D\left( t \right)$, which is due to the non-Hermiticity of the off-diagonal elements of ${H_0}\left( t \right)$, that is, $\Omega \left( t \right) \ne {g^ * }\left( t \right)$. However, if we assume that $h\left( t \right)$ and $e\left( t \right)$ are real functions, $A\left( t \right)$ will be equal to ${D^ * }(t)$, ${B_1}\left( t \right)$ will be purely imaginary and ${C_1}\left( t \right)$  will be real when we take the special case that $\Omega \left( t \right)$ is equal to ${g^ * }\left( t \right)$, which means ${H_1}\left( t \right)$ can return back to the one in Refs. \cite{Ibanez062136,Chen023405} by properly designing matrix elements. Furthermore, if we assume that $\Omega (t)$ and ${g }(t)$ both are real and $\Omega (t) = {g }(t)$, we have $A\left( t \right)=D\left( t \right)$, ${B_1}\left( t \right)=0$, and ${C_1}\left( t \right)$ is a complex function, where ${H_1}\left( t \right)$ returns back to the result in Ref. \cite{Ibanez023415} under proper matrix elements design.

In order to clarify essential difference between our results and previous results clearly, we  assume $\Omega \left( t \right){\rm{ = }}K\left( t \right){\rm{ + }}{J_{\rm{0}}}\left( t \right)$ and $g\left( t \right) = {K^ * }\left( t \right) + {J_1}\left( t \right)$, requiring $\left| {{J_{\rm{0}}}\left( t \right)} \right| \ll \left| {K\left( t \right)} \right|$ and $\left| {{J_1}\left( t \right)} \right| \ll \left| {K\left( t \right)} \right|$, where ${J_{\rm{0}}}\left( t \right)$ and ${J_1}\left( t \right)$ denote the perturbation of off-diagonal elements $\Omega \left( t \right)$ and $g\left( t \right)$, respectively. With the exact counteradiabatic driving Hamiltonian (\ref{H1_general}), we can obtain the approximate counteradiabatic driving Hamiltonian $H_{1}^\prime (t)$ as follows
\begin{widetext}
\scriptsize
\begin{equation}
\begin{split}
\begin{aligned}
H_{1}^\prime(t) =& \frac{{i\hbar }}{{4({{\left| K \right|}^2} + {d^2})}}\left( {\begin{array}{*{20}{c}}
{\dot K{K^*} - K{{\dot K}^*}}&{\dot K(e - h) - K(\dot e - \dot h)}\\
{{K^*}(\dot e - \dot h) - {{\dot K}^*}(e - h)}&{K{{\dot K}^*} - \dot K{K^*}}
\end{array}} \right) + \frac{{i\hbar }}{{4({{\left| K \right|}^2} + {d^2})}}&\\& \times \left( {\begin{array}{*{20}{c}}
{ - ({{\dot K}^*}{J_0} + K{{\dot J}_1} - {K^*}{{\dot J}_0} - \dot K{J_1} + {J_0}{{\dot J}_1} - {{\dot J}_0}{J_1})}&{{{\dot J}_0}(e - h) - {J_0}(\dot e - \dot h)}\\
{{J_1}(\dot e - \dot h) - {{\dot J}_1}(e - h)}&{{{\dot K}^*}{J_0} + K{{\dot J}_1} - {K^*}{{\dot J}_0} - \dot K{J_1} + {J_0}{{\dot J}_1} - {{\dot J}_0}{J_1}}
\end{array}} \right) - \frac{{i\hbar ({K^*}{J_0} + K{J_1} + {J_0}{J_1})}}{{4{{({{\left| K \right|}^2} + {d^2})}^2}}}&\\& \times \left( {\begin{array}{*{20}{c}}
{\dot K{K^*} - K{{\dot K}^*} - ({{\dot K}^*}{J_0} + K{{\dot J}_1} - {K^*}{{\dot J}_0} - \dot K{J_1} + {J_0}{{\dot J}_1} - {{\dot J}_0}{J_1})}&{(\dot K + {{\dot J}_0})(e - h) - (K + {J_0})(\dot e - \dot h)}\\
{({K^*} + {J_1})(\dot e - \dot h) - ({{\dot K}^*} + {{\dot J}_1})(e - h)}&{K{{\dot K}^*} - \dot K{K^*} + {{\dot K}^*}{J_0} + K{{\dot J}_1} - {K^*}{{\dot J}_0} - \dot K{J_1} + {J_0}{{\dot J}_1} - {{\dot J}_0}{J_1}}
\end{array}} \right),&
\label{H1_perturbation_original}
\end{aligned}
\end{split}
\end{equation}
\end{widetext}
whose detailed derivation process is shown in Appendix~\ref{APPC}. It should be noted that the quantities with superscript prime (e.g., $H_{1}^\prime(t)$ given by Eq.~(\ref{H1_perturbation_original}), ${\Omega_{a}^\prime}(t)$ and ${\Omega_{b}^\prime}(t)$ in Eq.~(\ref{Omega_aandbzz})) denote the approximate solution, while the ones without superscript prime (e.g., $H_{1}(t)$ given by Eq.~(\ref{H1_general}), ${\Omega_{a}}(t)$ and ${\Omega_{b}}(t)$ in Eq.~(\ref{tildeH1})) correspond to the exact expression. We show that Eq.~(\ref{H1_perturbation_original}) indicates the degree of deviation of the approximate counteradiabatic driving Hamiltonian (\ref{H1_perturbation_original}) from the exact counteradiabatic driving Hamiltonian (\ref{H1_general}). When $\left| {{J_{\rm{0}}}\left( t \right)} \right| \ll \left| {K\left( t \right)} \right|$ and $\left| {{J_1}\left( t \right)} \right| \ll \left| {K\left( t \right)} \right|$, Eq.~(\ref{H1_perturbation_original}) is almost consistent with Eq.~(\ref{H1_general}). With the increase of $\left| {{J_0}(t)} \right|$ and $\left| {{J_1}(t)} \right|$ compared with $\left| {K\left( t \right)} \right|$, Eq.~(\ref{H1_perturbation_original}) gradually deviates from Eq.~(\ref{H1_general}). The above conclusions can be reflected in Figs.~\ref{perturbation_exact_new} and \ref{perturbation_appr}, which are discussed in detail in Sec. III.

For further discussing the difference and relationship between our results and previous ones, we assume ${J_{\rm{0}}}\left( t \right)$ and ${J_1}\left( t \right)$ do not depend on the time and set ${J_{\rm{0}}}\left( t \right){\rm{ = }}{J_1}\left( t \right) \equiv J$, where $J$ is a time-independent complex number. Therefore, the off-diagonal elements $\Omega(t)$ and $g(t)$ can be easily rewritten as $\Omega \left( t \right) =K\left( t \right){\rm{ + }}J$ and $g\left( t \right) ={K^ * }\left( t \right) + J$.
Based on the assumptions, we consider three concrete cases as follows:

(i) For the first case, we take $J = 0$, which leads to $\Omega \left( t \right) = {g^ * }\left( t \right)$. Together with the setting $h(t) =  - \Delta (t)$ and $e(t) = \Delta (t)$, the
corresponding Hamiltonian ${H_0}(t)$ is the same as Refs. \cite{Ibanez062136,Chen023405}, where the exact counteradiabatic driving Hamiltonian ${H_1}\left( t \right)$ given by Eq.~(\ref{H1_general}) is also consistent with Refs.\cite{Ibanez062136,Chen023405}.

(ii) The second case is assumed as $J \ne 0$ and  ${\mathop{\rm Im}\nolimits} (J) \ne 0$, where we have $\Omega \left( t \right) \ne {g^ * }\left( t \right)$. In this case, we can expand Eq.~(\ref{H1_general}) in powers of $J$ as follows:
\begin{equation}
\begin{aligned}
{H_1}(t) = \sum\limits_{n = 0}^\infty  {{J^n}L^{(n)}(t)},
\label{H1expand1}
\end{aligned}
\end{equation}
which holds valid for $| J | < \min \{ | {{s_1}(t)} |,| {{s_2}(t)} |\} $ with ${s_1}(t) =  - {K_r}(t) + \sqrt {K_r^2(t) - [{{| {K(t)} |}^2} + {d^2}(t)]} $ and ${s_2}(t) =  - {K_r}(t) - \sqrt {K_r^2(t) - [{{| {K(t)} |}^2} + {d^2}(t)]}$ being the roots of ${C_1}(t)=0$, where ${K_r}(t) = \frac{1}{2}[K(t) + {K^ * }(t)]$ denotes the real part of $K(t)$. $L^{(n)}(t)$ in Eq.~(\ref{H1expand1}) is given by Eq.~(\ref{H1expandN}). The detailed derivation in Eq.~(\ref{H1expand1}) can be found in Appendix~\ref{APPD}.

Defining
\begin{equation}
\begin{aligned}
{c_N}(t) \equiv \sum\limits_{n = 0}^N {{J^n}L^{(n)}(t)},
\label{cN}
\end{aligned}
\end{equation}
if and only if $N$ tends to infinity and according to Eq.~(\ref{H1expand1}), we can get
\begin{equation}
\begin{aligned}
{H_1(t)} = \mathop {\lim }\limits_{N \to \infty } c_N(t)
\label{H1lim}
\end{aligned}
\end{equation}
which holds valid when $e(t)$, $h(t)$, $K(t)$ and $J$ are complex and $| J | < \min \{ | {{s_1}(t)} |,| {{s_2}(t)} |\} $.

(iii) For the third case, we assume that $J$ is a real number (i.e., ${\mathop{\rm Im}\nolimits} (J) = 0$)  and $J\ne 0$, where the off-diagonal elements $\Omega \left( t \right) =  K\left( t \right) + J$ and $g\left( t \right) = {K^ * }\left( t \right) + J$ are conjugate to each other, i.e., $\Omega \left( t \right) = {g^ * }\left( t \right)$, which means the corresponding Hamiltonian ${H_0}(t)$ is the same as Refs. \cite{Ibanez062136,Chen023405} when we set $h(t) =  - \Delta (t)$ and $e(t) = \Delta (t)$ with $\Delta (t)$ being a real function. We show that the derivation of Eq.~(\ref{H1lim}) does not make any approximations, which is completely equivalent to Eq.~(\ref{H1_general}). Therefore in this case, Eq.~(\ref{H1lim}) can return back to the previous results in Refs. \cite{Ibanez062136,Chen023405}.

However, when $N$ takes a finite integer but not equal to zero, we have $c_N(t) \ne {H_1}(t)$, that is to say, $c_N(t)$ given by Eq.~(\ref{cN}) cannot return back to the result of Eq.~(\ref{H1_general}). Especially, if we set $N=1$, we obtain the approximate counteradiabatic driving Hamiltonian
\begin{equation}
\begin{aligned}
{H_{1}^\prime}(t) = L^{(0)}(t) + JL^{(1)}(t)
\label{H1011z}
\end{aligned}
\end{equation}
when $| J | \ll \min \{ | {{s_1}(t)} |,| {{s_2}(t)} |\} $, where the specific forms of $L^{(0)}(t)$ and $L^{(1)}(t)$ are given by Eq.~(\ref{H1011}). In this case,  Eq.~(\ref{H1011z}) is no longer equivalent to Eq.~(\ref{H1_general}), which indicates the results of Refs. \cite{Ibanez062136,Chen023405} cannot be recovered when Eq.~(\ref{H1expand1}) (or Eq.~(\ref{H1lim})) is truncated to finite series ${c_N}(t)$ given by Eq.~(\ref{cN}), where $N$ takes a finite integer and $J$ is a real number ($J \ne 0$). For the case of setting $\Omega \left( t \right) =  K\left( t \right){\rm{ + }}J$ and $g\left( t \right) = {K^ * }\left( t \right) + J$ with $K\left( t \right) = {\Omega _R}\left( t \right)$, $h\left( t \right)=- \Delta \left( t \right)$, and $e\left( t \right)=\Delta \left( t \right) - i\Gamma $ \cite{Ibanez023415}, where ${\Omega _R}\left( t \right)$, $\Delta \left( t \right)$, $J$ and $\Gamma$ are real functions, we have similar discussions.

In summary, different from previous results, our result ${H_1(t)}$ in Eq.~(\ref{H1lim}) has the additional terms $\sum\nolimits_{n = 1}^\infty  {{J^n}L^{(n)}(t)} $ with $J$ being a complex number, which counts for corrections of the counteradiabatic driving Hamiltonian $L^{(0)}(t)$ with $J=0$. This can be explained by the fact that the introduction of perturbations $J$ leads to that the off-diagonal elements of Eq.~(\ref{general_H0}) are not conjugate to each other.

So far, we have derived the expressions of exact and approximate counteradiabatic driving Hamiltonians, which are given by Eqs.~(\ref{H1_general}) and (\ref{H1_perturbation_original}), respectively. From the above analysis, we can conclude that Eqs.~(\ref{general_H0}) and (\ref{H1_general}) are both Hamiltonians with general forms without considering any special conditions or perturbation approximation. That is to say, off-diagonal terms in the Hamiltonians (\ref{general_H0}) and (\ref{H1_general}) are not complex conjugate to each other and realizing such a Hamiltonian is usually a challenge in practice. However, recent experiments have shown that such ${H_0}\left( t \right)$ and ${H_1}\left( t \right)$ can be implemented in many physical systems in Refs.\cite{Chen548192,Wiersig063828,Wiersig012119,Wiersig203901,Peng123842,Wang111610,Lai57665,Holler1809}, which will be discussed in Sec. IV.

\section{Example of general two-level non-Hermitian system}

\subsection{Counteradiabatic driving Hamiltonian applied to a general two-level non-Hermitian system}
%figure1
\begin{figure}[t]
\centerline{
\includegraphics[width=8.8cm, height=4.8cm, clip]{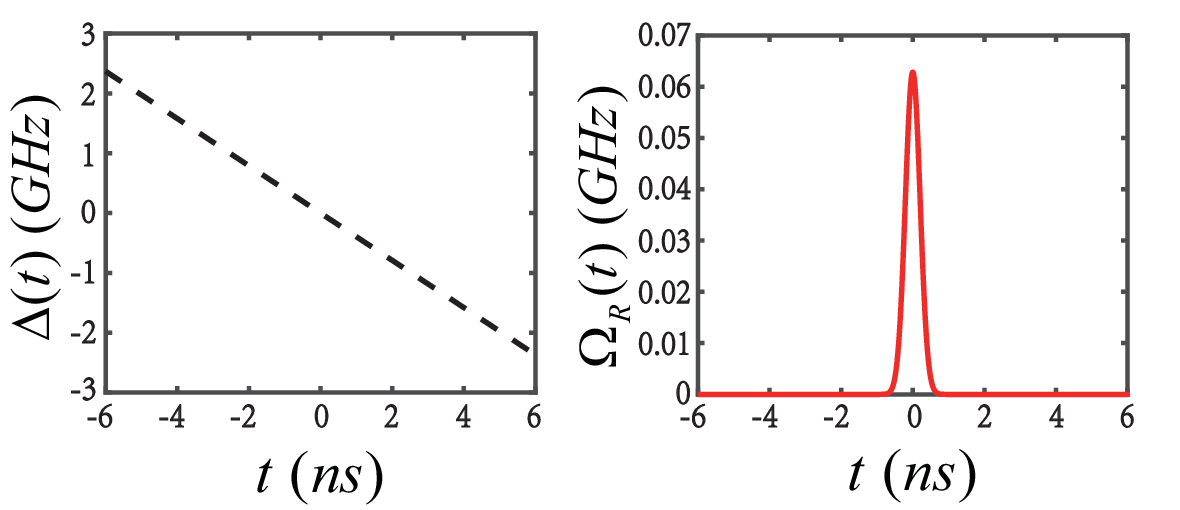}}
\caption{(Color online) $\Delta (t) =  - 2{y_0}t$ (black dashed line) and ${\Omega _R}(t) = {\Omega _0}{e^{ - {x_0}{t^2}}}$ (red solid line) for a Gaussian pulse in Eq.~(\ref{tilde_H0_model}). The parameters chosen are ${x_0} = {(2\pi )^2} \times 0.3$ GHz$^2$, ${y_0} = {(2\pi )^2} \times 0.005$ GHz$^2$, and ${\Omega _0} = 2\pi  \times 0.01$ GHz.
} \label{pulses_t_new}
\end{figure}

%figure2
\begin{figure*}
\centerline{
\includegraphics[width=18cm, height=10.8cm, clip]{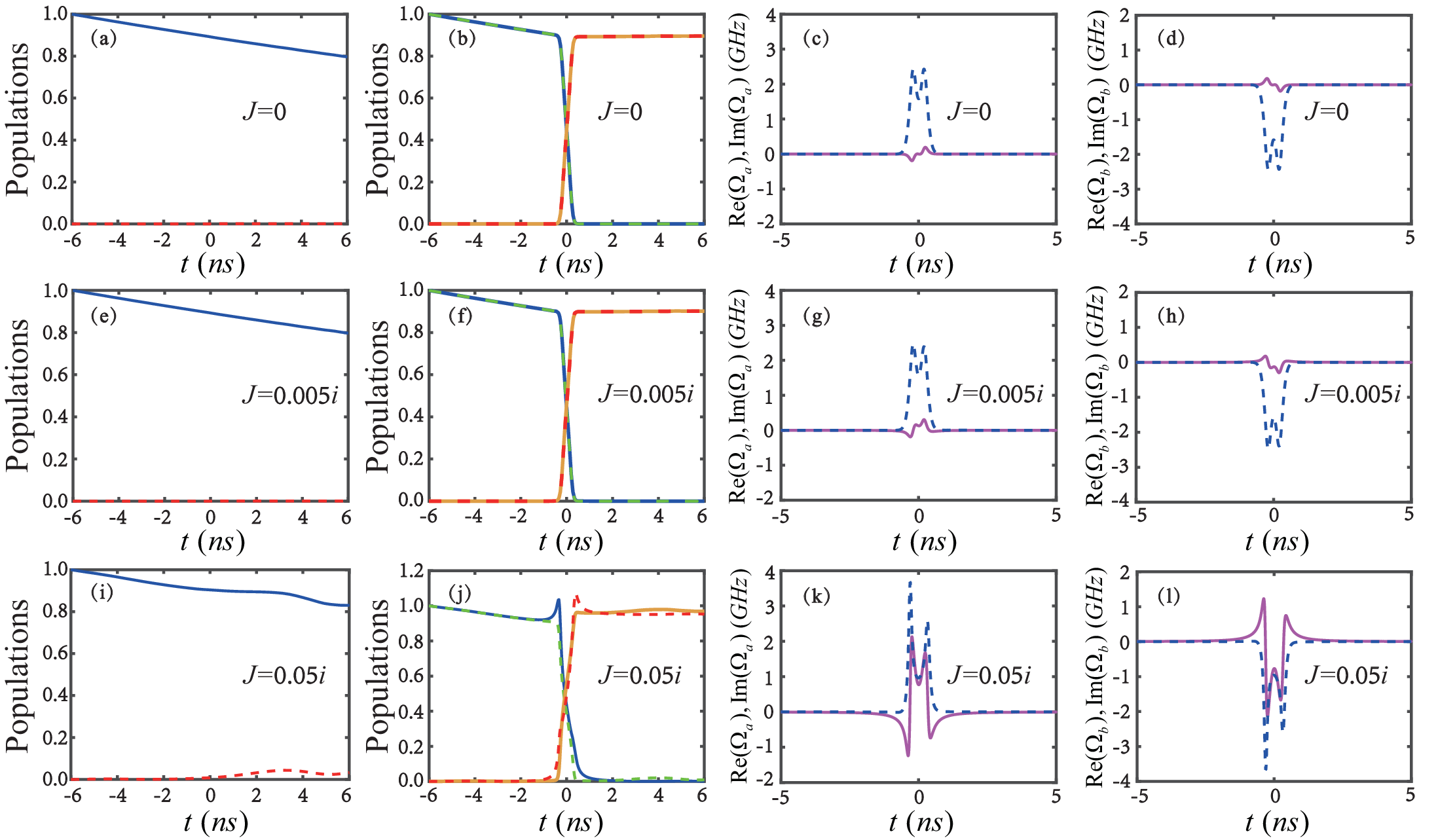}}
\caption{(Color online) The first column shows populations of the ground state ${P_1}(t)$ (red dashed lines) and of the excited state ${P_2}(t)$ (blue solid lines) controlled by ${H_{0}}(t)$ given by Eq.~(\ref{tilde_H0_model}). The second column shows the populations under the control of two kinds of Hamiltonians: ${P_1}(t)$ (red dashed lines) and ${P_2}(t)$ (blue solid lines) controlled by exact total Hamiltonian ${{H}}\left( t \right) = {{H}_{0}}\left( t \right) + {{H}_{1}}\left( t \right)$ with ${{H}_{0}}\left( t \right)$ and ${{H}_{1}}\left( t \right)$ respectively given by Eqs.~(\ref{tilde_H0_model}) and (\ref{tildeH1}); ${P_1}(t)$ (orange solid lines) and ${P_2}(t)$ (green dashed lines) controlled by ${H}(t)$ which is approximated by neglecting ${\mathop{\rm Re}\nolimits} [{\Omega _a}(t)]$ and ${\mathop{\rm Re}\nolimits} [{\Omega _b}(t)]$ in ${{H}_{1}}(t)$. The third and fourth column show the real (purple solid lines) and imaginary parts (blue dashed lines) of the off-diagonal terms in Eq.~(\ref{tildeH1}) which is the exact counteradiabatic driving Hamiltonian. Different rows correspond to different cases of $J = 0$ GHz for (a)-(d), $J = 0.005i$ GHz for (e)-(h), and $J = 0.05i$ GHz for (i)-(l), respectively. The parameters are chosen as ${\gamma _1} = 2\pi  \times 0.1$ MHz, ${\gamma _2} = 2\pi  \times 3$ MHz, and ${\omega _L} = 0.005\pi $ GHz. Other parameters are the same as Fig.~\ref{pulses_t_new}.}\label{perturbation_exact_new}
\end{figure*}

%figure3
\begin{figure*}
\centerline{
\includegraphics[width=18cm, height=10.8cm, clip]{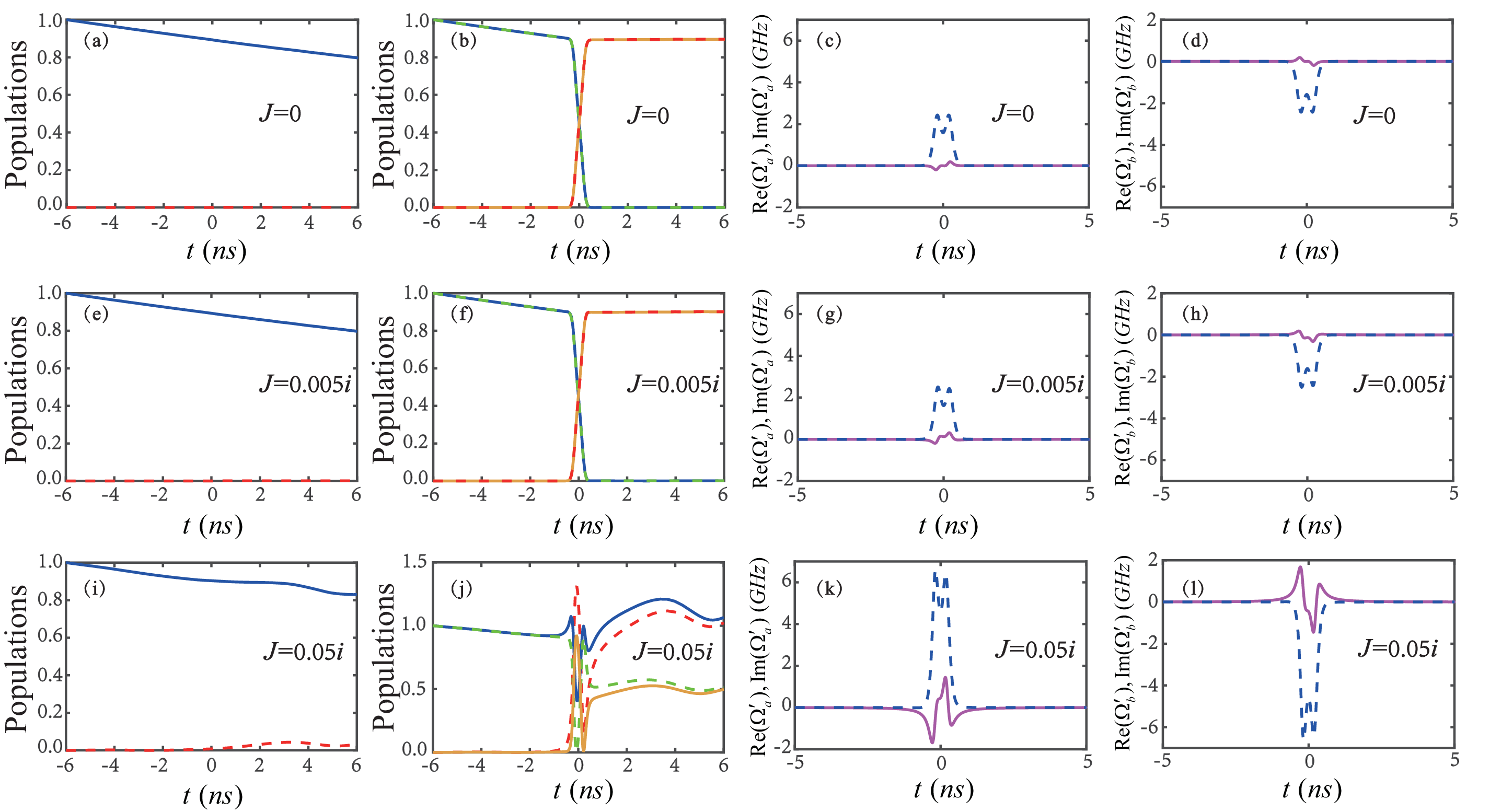}}
\caption{(Color online) The first column shows ${P_1}(t)$ (red dashed lines) and ${P_2}(t)$ (blue solid lines) controlled by ${H_{0}}(t)$ given by Eq.~(\ref{tilde_H0_model}). The second column shows the populations under the control of two kinds of Hamiltonians: ${P_1}(t)$ (red dashed lines) and ${P_2}(t)$ (blue solid lines) controlled by approximate total Hamiltonian $H^\prime (t){\rm{ = }}{H_{0}}(t) + H_{1}^\prime (t)$ with ${{H}_{0}}\left( t \right)$ and $H_{1}^\prime (t)$ respectively given by Eqs.~(\ref{tilde_H0_model}) and (\ref{Omega_aandbzz}); ${P_1}(t)$ (orange solid lines) and ${P_2}(t)$ (green dashed lines) controlled by $H^\prime (t)$ which is approximated by neglecting ${\mathop{\rm Re}\nolimits} [\Omega _a^\prime (t)]$ and ${\mathop{\rm Re}\nolimits} [\Omega _b^\prime (t)]$ in $H_{1}^\prime (t)$ given by Eq.~(\ref{Omega_aandbzz}). The third and fourth column show the real (purple solid lines) and imaginary parts (blue dashed lines) of the off-diagonal terms for approximate counteradiabatic driving Hamiltonian $H_{1}^\prime (t)$ given by Eq.~(\ref{Omega_aandbzz}). Different rows denote different cases of $J = 0$ GHz for (a)-(d), $J = 0.005i$ GHz for (e)-(h), and $J = 0.05i$ GHz for (i)-(l), respectively. The parameters are as described in Fig.~\ref{perturbation_exact_new}.}\label{perturbation_appr}
\end{figure*}

%figure4
\begin{figure}
\centerline{
\includegraphics[width=8cm, height=5.8cm, clip]{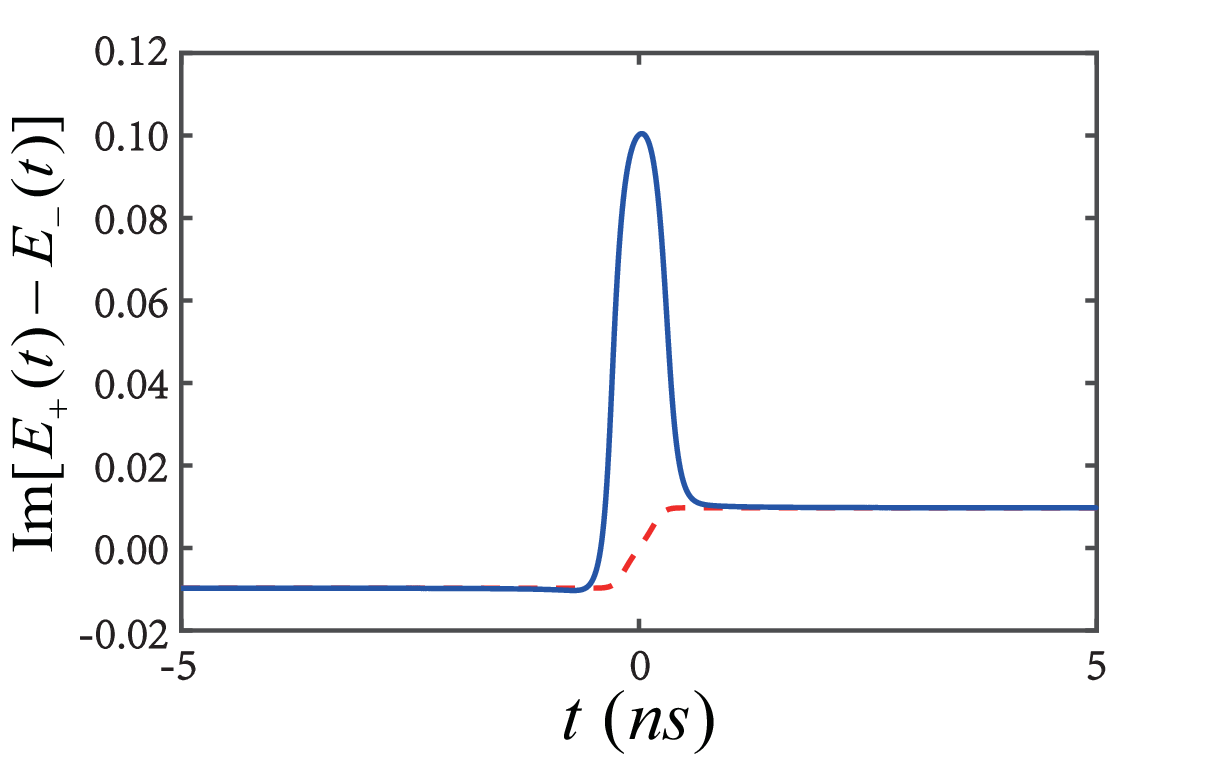}}
\caption{(Color online) Differences of the imaginary parts of the instantaneous eigenvalues of ${H_{0}}(t)$ in Eq.~(\ref{tilde_H0_model}). The parameters are chosen as $J = 0$ GHz (red dashed line) and $J = 0.05i$ GHz (blue solid line). Other parameters are chosen as Fig.~\ref{pulses_t_new}.}\label{eigenvalue}
\end{figure}

In the following, we take a two-level non-Hermitian system as an example to display the feasibility of the idea proposed above. As an application of the general approach of the previous section, we shall speed up adiabatic processes in the system under study. We assume that a two-level non-Hermitian system has a ground level $\left| 1 \right\rangle  = {\left( {\begin{array}{*{20}{c}}
1&0\end{array}} \right)^T}$ and an excited level $\left| 2 \right\rangle  = {\left( {\begin{array}{*{20}{c}}
0&1\end{array}} \right)^T}$. A particular case of practical importance is population inversion, which generalizes the shortcut techniques described for a two-level non-Hermitian system. We shall  assume a semiclassical treatment of the interaction between a laser electric field linearly polarized and gain rate ${\gamma _1}\left( t \right)$, loss rate ${\gamma _2}\left( t \right)$ from the ground state and excited state, respectively. Applying the electric dipole approximation and a laser-adapted interaction picture, the Hamiltonian with disregarding atomic motion \cite{Peng123842,Wang111610,Lai57665,Holler1809,Ibanez013428} is
\begin{eqnarray}
\begin{aligned}
{H_0}(t) = \hbar \left( {\begin{array}{*{20}{c}}
{\frac{1}{2}[ - \Delta (t) + i{\gamma _1}(t)]}&{\Omega (t)}\\
{g(t)}&{\frac{1}{2}[\Delta (t) - i{\gamma _2}(t)]}
\end{array}} \right),
\label{tilde_H0_model}
\end{aligned}
\end{eqnarray}
where $\Omega (t) = \frac{1}{2}{\Omega _R}(t)(1 + {e^{ - 2i{\omega _L}t}}) +  J$, $g(t) = \frac{1}{2}{\Omega _R}(t)(1 + {e^{2i{\omega _L}t}}) + J$ and $\Delta (t) = {\omega _0}(t) - {\omega _L}$ with $J$ denoting a time-independent complex number. ${\Omega _R}(t)$  is the Rabi frequency. ${\Delta (t)}$ is the detuning from the atomic time-dependent transition frequency ${\omega _0}$. ${\omega _L}$ is the frequency of driving field. In order to further study and demonstrate the influence of pulse perturbation $J$ on the atomic dynamics, we shall explore the difference of numerical simulation results in population transfer when $J$ takes different values here. With Eq.~(\ref{H1_general}) by defining ${\Omega _1}(t) = {\Omega _R}(t)(1 + {e^{ - 2i{\omega _L}t}}) + 2J$ and ${\Omega _2}(t) = {\Omega _R}(t)(1 + {e^{2i{\omega _L}t}}) + 2J$, we find that the corresponding exact counteradiabatic driving Hamiltonian ${{H}_{1}}(t)$ can be written as
\begin{widetext}
\begin{equation}
\begin{aligned}
{H_{1}}(t) = \left( {\begin{array}{*{20}{c}}
{\nu (t) [{{\dot \Omega }_1}(t){\Omega _2}(t) - {\Omega _1}(t){{\dot \Omega }_2}(t)]}&{{\Omega _a}(t)}\\
{{\Omega _b}(t)}&{\nu (t) [{\Omega _1}(t){{\dot \Omega }_2}(t) - {{\dot \Omega }_1}(t){\Omega _2}(t)]}
\end{array}} \right),
\label{tildeH1}
\end{aligned}
\end{equation}
\end{widetext}
where
\begin{equation}
\begin{aligned}
{\rm{ }}{\Omega _a}(t) ={} &\nu (t)\{ {\dot \Omega _1}(t)\{ 2\Delta (t) - i[{\gamma _1}(t) + {\gamma _2}(t)]\}  &\\
&- {\Omega _1}(t)\{ 2\dot \Delta (t) - i[{\dot \gamma _1}(t) + {\dot \gamma _2}(t)]\} \},\\
{\Omega _b}(t) ={} &\nu (t)\{ {\Omega _2}(t)\{ 2\dot \Delta (t) - i[{{\dot \gamma }_1}(t) + {{\dot \gamma }_2}(t)]\}  &\\
&- {{\dot \Omega }_2}(t)\{ 2\Delta (t) - i[{\gamma _1}(t) + {\gamma _2}(t)]\} \}
\label{Omega_aandb}
\end{aligned}
\end{equation}
with $\nu (t)  = \frac{{i\hbar }}{{{{\{ 2\Delta (t) - i[{\gamma _1}(t) + {\gamma _2}(t)]\} }^2} + 4{\Omega _1}(t){\Omega _2}(t)}}$.
Eq.~(\ref{tildeH1}) sets the transitionless shortcut protocol, in which
%With this protocol, population inversion is always achieved even if the adiabatic condition fails for ${H_{0}}(t)$.
the new engineered Hamiltonian ${{H}}\left( t \right) = {{H}_{0}}\left( t \right) + {{H}_{1}}\left( t \right)$ can drive the adiabatic states along the adiabatic path but without generating any transition probabilities between them, which is governed by Schr${\rm\ddot o}$dinger equation $i\hbar | {\dot \Psi (t)} \rangle  = {H}\left( t \right)\left| {\Psi (t)} \right\rangle$ with $|\Psi(t) \rangle$ given by Eq.~(\ref{solution_final}) and does not undergo adiabatic process. With Eq.~(\ref{tilde_H0_model}), we can  obtain the corresponding approximate Hamiltonian
\begin{equation}
\begin{aligned}
{H'_1}(t){\rm{ = }}\left( {\begin{array}{*{20}{c}}
{{\Delta_{1}^\prime}(t)}&{{\Omega_{a}^\prime}(t)}\\
{{\Omega_{b}^\prime}(t)}&{{\Delta_{2}^\prime}(t)}
\end{array}} \right),
\label{Omega_aandbzz}
\end{aligned}
\end{equation}
whose matrix elements are respectively given by Eq.~(\ref{H1_perturbation_original}) with the replacements by $h(t)=\frac{1}{2}[ - \Delta (t) + i{\gamma _1}(t)]$, $e(t)=\frac{1}{2}[\Delta (t) - i{\gamma _2}(t)]$, $K (t) = \frac{1}{2}{\Omega _R}(t)(1 + {e^{ - 2i{\omega _L}t}})$, and ${J_0}(t)={J_1}(t) \equiv J$. In the following, we will apply the general Hamiltonian ${{H}_{0}}\left( t \right)$ in Eq.~(\ref{tilde_H0_model}) and discuss the population transfer under the control of the corresponding exact total Hamiltonian ${{H}}\left( t \right) = {{H}_{0}}\left( t \right) + {{H}_{1}}\left( t \right)$ with ${{H}_{1}}\left( t \right)$ in Eq.~(\ref{tildeH1}) and approximate total Hamiltonian $H^\prime (t)= {{H}_{0}}\left( t \right) + {H_{1}^\prime}(t)$ with ${H_{1}^\prime}(t)$ given by Eq.~(\ref{Omega_aandbzz}), respectively, in two cases, i.e., $J = 0$ GHz and $J \ne 0$ GHz.

We now study the forced population transfer by considering a linearly chirped Gaussian pulse with detuning $\Delta (t) =  - 2{y_0}t$ and Gaussian Rabi frequency ${\Omega _R}(t) = {\Omega _0}{e^{ - {x_0}{t^2}}}$ in Eq.~(\ref{tilde_H0_model}) with parameters ${x_0} = {(2\pi )^2} \times 0.3$ GHz$^2$, ${y_0} = {(2\pi )^2} \times 0.005$ GHz$^2$, and ${\Omega _0} = 2\pi  \times 0.01$ GHz, as shown in Fig.~\ref{pulses_t_new}. For convenience, we take ${\gamma _1}$ and ${\gamma _2}$ as constants, although they could also depend on time in a general case as effective gain and loss rate controlled by further interactions. In Fig.~\ref{perturbation_exact_new}(a)(e)(i), we show that ${{H}_{0}}(t)$ given by Eq.~(\ref{tilde_H0_model}) does not invert the populations of ${P_1}(t)$ and ${P_2}(t)$ under the used current parameters whether $J = 0$ GHz or $J \ne 0$ GHz ($J=0.005i$ GHz and $J=0.05i$ GHz), where the initial conditions for the populations of the ground and excited states are ${P_1}\left( 0 \right) = 0$ and ${P_2}\left( 0 \right) = 1$. In order to achieve efficient population inversion in a short time, it is necessary to introduce shortcuts to adiabaticity. ${{H}_{1}}(t)$ in Eq.~(\ref{tildeH1}) can be used as the exact counteradiabatic driving Hamiltonian to accelerate the adiabatic dynamics governed by the reference Hamiltonian ${{H}_{0}}(t)$ in Eq.~(\ref{tilde_H0_model}), which has off-diagonal terms with real and imaginary parts depicted in Fig.~\ref{perturbation_exact_new}(c) and \ref{perturbation_exact_new}(d) in the case of $J = 0$ GHz, where the imaginary parts can be realized by a complementary laser with orthogonal polarization, while the real parts constitute a non-Hermitian contribution. We show that the processes of adiabatic quantum state transfer when $J = 0$ GHz for the reference Hamiltonian ${{H}_{0}}(t)$ and for the total Hamiltonian ${{H}}\left( t \right) = {{H}_{0}}\left( t \right) + {{H}_{1}}\left( t \right)$ with ${{H}_{1}}\left( t \right)$ given by Eq.~(\ref{tildeH1}) correspond to Fig.~\ref{perturbation_exact_new}(a) and \ref{perturbation_exact_new}(b), respectively.

Now we explore the influence of non-Hermitian off-diagonal terms of the total Hamiltonian on population transfer, which is as shown in Fig.~\ref{perturbation_exact_new}(b), i.e., ${P_1}(t)$ (orange solid line) and ${P_2}(t)$ (green dashed line) with the real parts of ${\Omega _a}(t)$ and ${\Omega _b}(t)$ being ignored. We can see that these two types of the population transfer (one is under the control of total Hamiltonian ${{H}}\left( t \right) = {{H}_{0}}\left( t \right) + {{H}_{1}}\left( t \right)$, the other is under the control of ${{H}}\left( t \right)$ when the real parts of ${\Omega _a}(t)$ and ${\Omega _b}(t)$ in ${{H}_{1}}\left( t \right)$ are ignored) shown in Fig.~\ref{perturbation_exact_new}(b) are almost the same, which indicates the real parts of ${\Omega _a}(t)$ and ${\Omega _b}(t)$ contribute very small to the dynamics. This can be shown in purple solid lines of Fig.~\ref{perturbation_exact_new}(c) and \ref{perturbation_exact_new}(d), where ${\mathop{\rm Re}\nolimits} [{\Omega _a}(t)]$ and ${\mathop{\rm Re}\nolimits} [{\Omega _b}(t)]$ are much smaller than ${\mathop{\rm Im}\nolimits} [{\Omega _a}(t)]$ and ${\mathop{\rm Im}\nolimits} [{\Omega _b}(t)]$. Therefore, the real parts of ${\Omega _a}(t)$ and ${\Omega _b}(t)$ (i.e., ${\mathop{\rm Re}\nolimits} [{\Omega _a}(t)]$ and ${\mathop{\rm Re}\nolimits} [{\Omega _b}(t)]$) can be ignored in this case. Similar to Ref.\cite{Ibanez023415}, the off-diagonal elements of Eq.~(\ref{tildeH1}) can be approximately regarded as Hermitian elements at this case, which means that it is weak non-Hermiticity regime when $J = 0$ GHz for both ${{H}_{0}}\left( t \right)$ and ${{H}_{1}}\left( t \right)$ given by Eqs.~(\ref{tilde_H0_model}) and (\ref{tildeH1}). Therefore the Hamiltonian in Eq.~(\ref{tildeH1}) can be realized approximately. When the real parts for the off-diagonal elements of Eq.~(\ref{tildeH1}) can not be ignored, the realization mentioned above is not valid anymore.

However, with the increase of intensity of pulse perturbation $J$, the non-Hermiticity of ${{H}_{0}}(t)$ in Eq.~(\ref{tilde_H0_model}) and the corresponding exact counteradiabatic driving Hamiltonian ${{H}_{1}}(t)$ in Eq.~(\ref{tildeH1}) becomes strong and the above conclusion will not be available. Here, we consider the case where the order of magnitude of $| J |$ is small relative to $|\frac{1}{2}{\Omega _R}(t)(1 + {e^{ - 2i{\omega _L}t}})|$ (see Fig.~\ref{perturbation_exact_new}(e)(f)(g)(h)). Compared with the imaginary parts  of ${\Omega _a}(t)$ and ${\Omega _b}(t)$, we can see that the real parts of ${\Omega _a}(t)$ and ${\Omega _b}(t)$  have a small contribution to the dynamics when $| J |$ is small enough (see Fig.~\ref{perturbation_exact_new}(g) and \ref{perturbation_exact_new}(h)). However, a large contribution occurs when $| J |$ increases to the same order of magnitude as $|\frac{1}{2}{\Omega _R}(t)(1 + {e^{ - 2i{\omega _L}t}})|$ (see Fig.~\ref{perturbation_exact_new}(k) and \ref{perturbation_exact_new}(l)). Therefore, the population transfer in the weak non-Hermiticity regime controlled by total Hamiltonian ${{H}}(t) = {{H}_{0}}(t) + {{H}_{1}}(t)$ with ${{H}_{0}}\left( t \right)$ and ${{H}_{1}}\left( t \right)$ respectively given by Eqs.~(\ref{tilde_H0_model}) and (\ref{tildeH1}) neglecting the real parts of ${\Omega _a}(t)$ and ${\Omega _b}(t)$ in ${H_{1}}(t)$ is in good agreement with that controlled by total Hamiltonian ${H}(t)$, but the one in the strong non-Hermiticity regime is not, which can be found in Fig.~\ref{perturbation_exact_new}(f) and \ref{perturbation_exact_new}(j), respectively.

We show that Fig.~\ref{perturbation_exact_new} corresponds to the situation under the control of the exact counteradiabatic driving Hamiltonian (\ref{tildeH1}) while Fig.~\ref{perturbation_appr} corresponds to the situation under the control of the approximate counteradiabatic driving Hamiltonian $H_{1}^\prime (t)$ given by Eq.~(\ref{Omega_aandbzz}). Fig.~\ref{perturbation_appr}(a)(e)(i) are similar to those in Fig.~\ref{perturbation_exact_new}, which denote the population transfers driven by the original Hamiltonian (\ref{tilde_H0_model}) without considering $H_{1}^\prime (t)$. The approximate counteradiabatic driving Hamiltonian (\ref{Omega_aandbzz}) also has off-diagonal terms with real and imaginary parts, see Fig.~\ref{perturbation_appr}(c)(d), (g)(h) and (k)(l). Fig.~\ref{perturbation_appr}(b)(f)(j) describe the population of the ground state ${P_1}(t)$ (red dashed lines) and of the excited state ${P_2}(t)$ (blue solid lines) for the total Hamiltonian $H^\prime (t){\rm{ = }}{H_{0}}(t) + H_{1}^\prime (t)$ with ${{H}_{0}}\left( t \right)$ and $H_{1}^\prime (t)$ respectively given by Eqs.~(\ref{tilde_H0_model}) and (\ref{Omega_aandbzz}), together with the populations ${P_1}(t)$ (orange solid lines) and ${P_2}(t)$ (green dashed lines) when the total Hamiltonian $H^\prime (t)$ is approximated by neglecting ${\mathop{\rm Re}\nolimits} [\Omega _a^\prime (t)]$ and ${\mathop{\rm Re}\nolimits} [\Omega _b^\prime (t)]$ in Eq.~(\ref{Omega_aandbzz}). We show that the Hermiticity of the off-diagonal of total approximate Hamiltonian $H^\prime (t)$ is almost satisfied when $J=0$ GHz, see Fig.~\ref{perturbation_appr}(b)(c)(d).

However, as $J$ increases to $0.05i$ GHz, the off-diagonal elements of $H^\prime _1(t)$ reaches the non-Hermitian region, which is shown in Fig.~\ref{perturbation_appr}(j)(k)(l). Namely, $J$ reflects the non-Hermiticity of the off-diagonal elements of Hamiltonians (both Eqs.~(\ref{tilde_H0_model}) and (\ref{Omega_aandbzz})) to some extent. Specifically, the population inversion under the control of total approximate Hamiltonian $H^\prime (t){\rm{ = }}{H_{0}}(t) + H_{1}^\prime (t)$ with ${\mathop{\rm Re}\nolimits} [\Omega _a^\prime (t)]$ and ${\mathop{\rm Re}\nolimits} [\Omega _b^\prime (t)]$ of the approximate counteradiabatic driving Hamiltonian $H_{1}^\prime (t)$ being ignored is in good agreement with that controlled by total approximate Hamiltonian $H^\prime (t)$ in the weak non-Hermiticity regime (see Fig.~\ref{perturbation_appr}(b) and \ref{perturbation_appr}(f)), but the one in the strong non-Hermiticity regime is broken (see Fig.~\ref{perturbation_appr}(j)). Moreover, the population inversion under the control of $H^\prime (t){\rm{ = }}{H_{0}}(t) + H_{1}^\prime (t)$ can be realized when $| J |$ is equal or close to zero (see Fig.~\ref{perturbation_appr}(b) and \ref{perturbation_appr}(f)) while it fails when $| J |$ increases to some certain value (see Fig.~\ref{perturbation_appr}(j)). This is reasonable because the approximate Hamiltonian (\ref{H1_perturbation_original}) indicates its deviation from the exact counteradiabatic driving Hamiltonian (\ref{H1_general}), where Eq.~(\ref{H1_perturbation_original}) is almost equivalent to Eq.~(\ref{H1_general}) when $\left| J \right| \ll \left| K(t) \right|$ (see Fig.~\ref{perturbation_exact_new}(a)(b)(c)(d) and Fig.~\ref{perturbation_appr}(a)(b)(c)(d)), while it has a very small deviation from Eq.~(\ref{H1_general}) with the increase of $|J|$ (see Fig.~\ref{perturbation_exact_new}(e)(f)(g)(h) and Fig.~\ref{perturbation_appr}(e)(f)(g)(h)). With the further increase of $|J|$ compared with $\left| K(t) \right|$ (see Fig.~\ref{perturbation_exact_new}(i)(j)(k)(l) and Fig.~\ref{perturbation_appr}(i)(j)(k)(l)), Eq.~(\ref{H1_perturbation_original}) is not available, because the tenable condition of Eq.~(\ref{H1_perturbation_original}) is $\left| J \right| \ll \left| K(t) \right|$, which leads to the failure of population inversion (see the blue solid and red dashed lines or the orange solid and green dashed lines in Fig.~\ref{perturbation_appr}(j)). That is to say, the expression of Eq.~(\ref{H1_perturbation_original}) is valid only in the case of weak non-Hermiticity. When $| J |$ increases to the same order of magnitude as $| K(t) |$, i.e., $|\frac{1}{2}{\Omega _R}(t)(1 + {e^{ - 2i{\omega _L}t}})|$ here, this result is no longer available. However, the general exact solution (\ref{H1_general}) we derived before is always available whether in the weak non-Hermiticity regime or in the strong non-Hermiticity regime. We show that there is another explanation for the weak non-Hermiticity of ${{H}_{0}}\left( t \right)$ in Fig.~\ref{eigenvalue}, which is the difference values between the imaginary parts of the two eigenvalues of ${{H}_{0}}\left( t \right)$ for different $J$. It is obvious that ${\mathop{\rm Im}\nolimits} [{E_ + }(t) - {E_ - }(t)]$ with $J = 0$ GHz is closer to zero. That is to say, $\exp \{  - \int_0^t {{\mathop{\rm Im}\nolimits} [{E_ + }(t) - {E_ - }(t)]dt} \}  \approx 1$ when $J = 0$ GHz, which indicates a weak non-Hermiticity regime \cite{Ibanez023415}, while $J = 0.05i$ GHz corresponds to the strong non-Hermiticity regime. So far, we have extended the transitionless driving algorithm to the strong non-Hermiticity regime.

From Fig.~\ref{eigenvalue}, we can also see that the imaginary parts of the two eigenvalues are not zero. Thus, in the complex energy space, the two energy level can not be crossing each other due to the occurring of the imaginary eigenvalues. The existence of the imaginary eigenvalues means that the energy and populations of the system are all non-conservative in the process, which is the general characteristics of the non-Hermitian systems without PT-symmetry.

\subsection{Transitionless driving scheme applied to a general two-level non-Hermitian system beyond the RWA}
%figure5
\begin{figure}[b]
\centerline{
\includegraphics[width=8cm, height=12.8cm, clip]{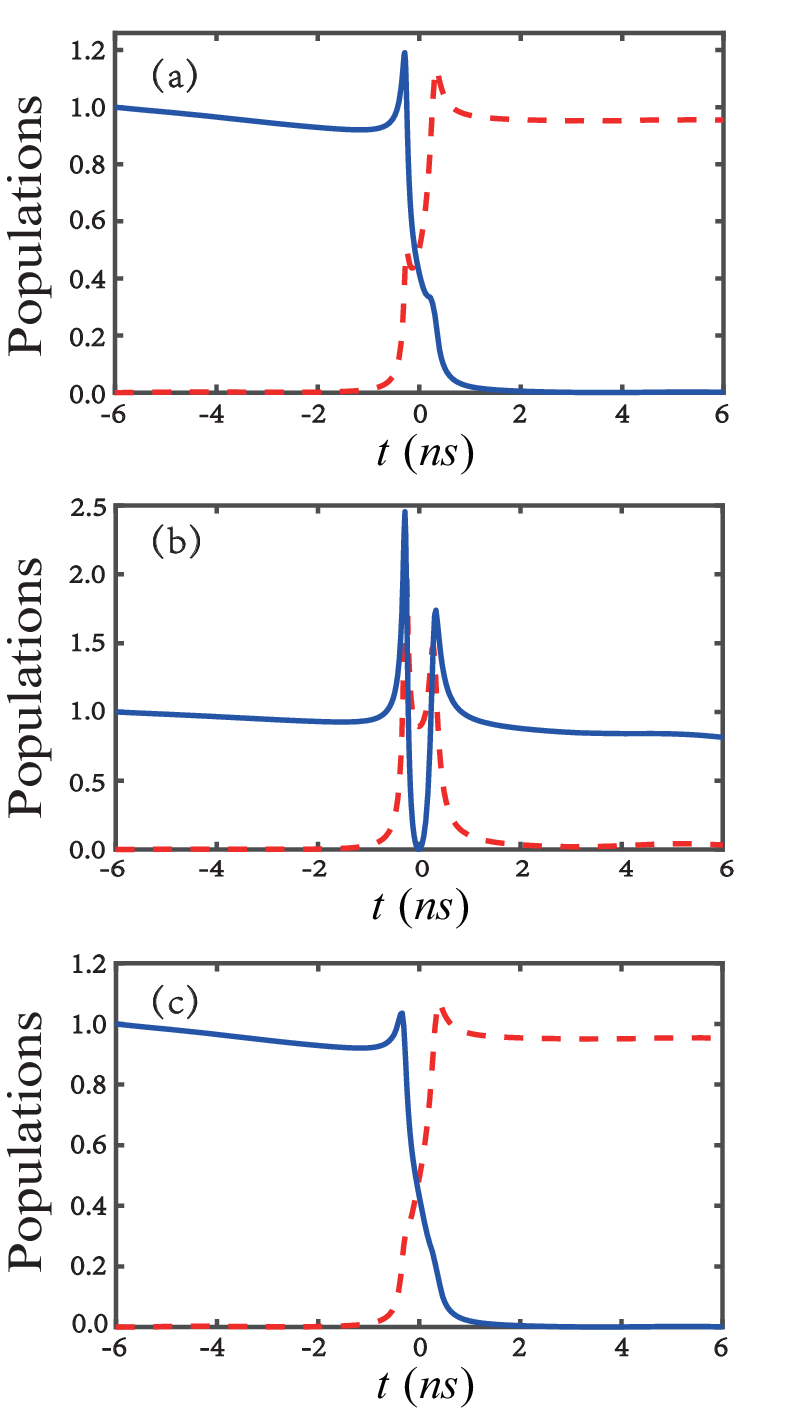}}
\caption{(Color online) The evolution of the population for a two-level system with gain rate ${\gamma _1} = 2\pi  \times 0.1$ MHz and loss rate ${\gamma _2} = 2\pi  \times 3$ MHz in Eq.~(\ref{H0_model_RWA}). Here we choose $J = 0.05i$ GHz. We consider a linearly chirped Gaussian pulse with detuning $\Delta (t) =  - 2{y_1}t$ and Gaussian Rabi frequency ${\Omega _R}(t) = {\Omega _0}{e^{ - {x_1}{t^2}}}$ with parameters ${x_1} = {(2\pi )^2} \times 0.3$ GHz$^2$, ${y_1} = {(2\pi )^2} \times 0.005$ GHz$^2$. The other parameters: ${\Omega _0} = 2\pi  \times 0.01$ GHz, ${\omega _L} = 0.005\pi $ GHz. ${P_1}\left( t \right)$ (red dashed lines) and ${P_2}\left( t \right)$ (blue solid lines) are shown. The population transfer is implemented by the shortcuts to adiabaticity technique under the RWA [Fig.~\ref{population_CR_new}(a) and \ref{population_CR_new}(b)] and beyond the RWA [Fig.~\ref{population_CR_new}(c)]. Note that all the counter-rotating terms are neglected in Fig.~\ref{population_CR_new}(a), but are considered in Fig.~\ref{population_CR_new}(b).}\label{population_CR_new}
\end{figure}

%figure6
\begin{figure}[t]
\centerline{
\includegraphics[width=6cm, height=13.8cm, clip]{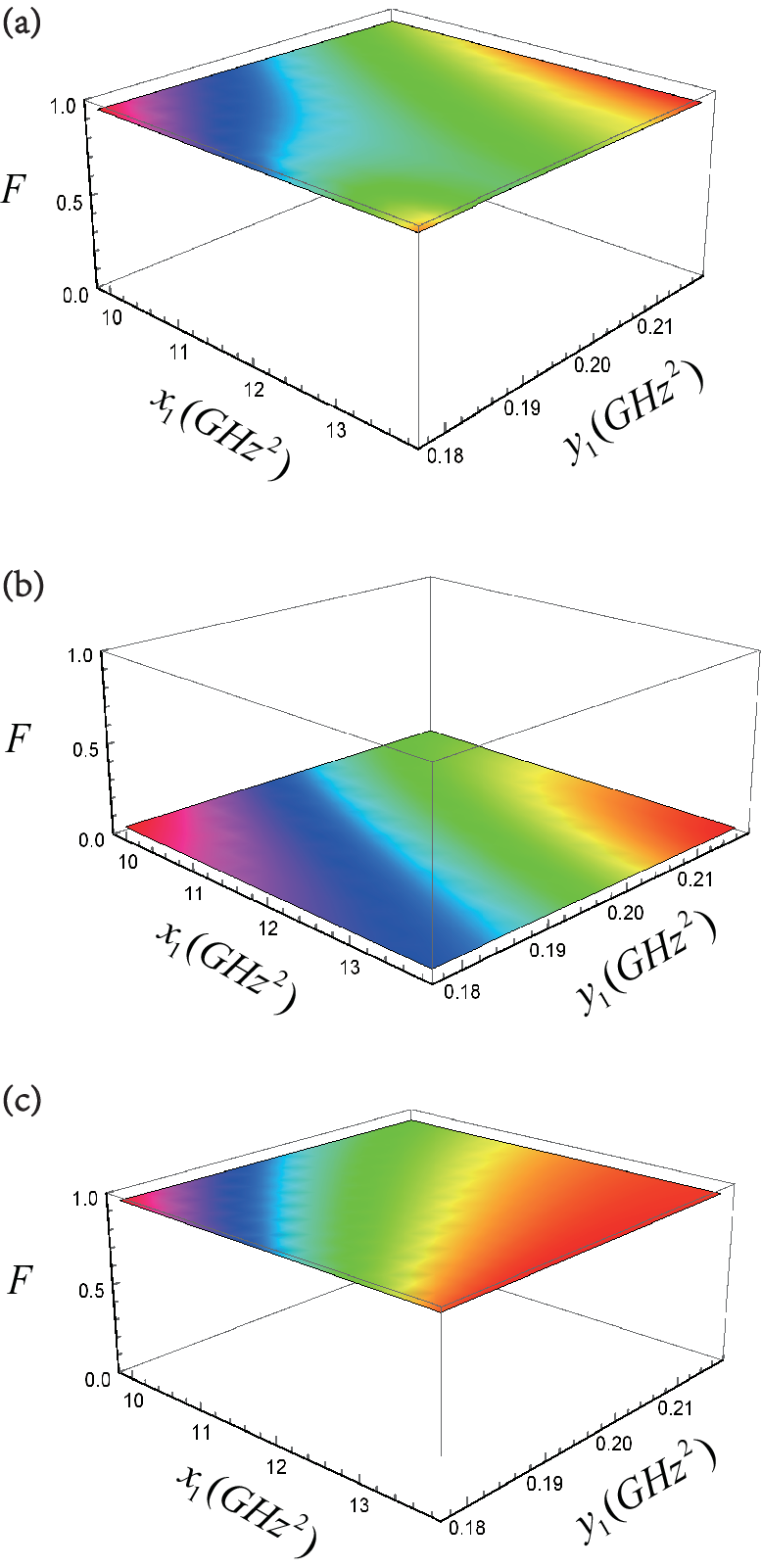}}
\caption{Fidelity between the finial state $| {{\psi _{{t_f}}}}\rangle $ and the target state $| 1 \rangle $ as a function of ${x_1}$ and ${y_1}$ with ${t_f}=6$ ns. The population transfer is implemented by the shortcuts to adiabaticity technique under the RWA [Fig.~\ref{fidelity_CR_new}(a) and \ref{fidelity_CR_new}(b)] and implemented by the shortcuts to adiabaticity technique beyond the RWA [Fig.~\ref{fidelity_CR_new}(c)]. We show that all the counter-rotating terms are neglected in Fig.~\ref{fidelity_CR_new}(a), but they are considered in Fig.~\ref{fidelity_CR_new}(b). The parameters are chosen as ${x_1} = {(2\pi )^2} \times [0.25,0.35]$ GHz$^2$ and ${y_1} = {(2\pi )^2} \times [0.0045,0.0055]$ GHz$^2$. Other parameters are chosen as Fig.~\ref{population_CR_new}.}\label{fidelity_CR_new}
\end{figure}
In this section, we present a shortcuts to adiabaticity technique for an open two-level quantum system beyond the RWA in the strong non-Hermiticity regime. The RWA is in common used in atomic optics and magnetic resonance \cite{Irish173601,Chan065507}, while it is only valid in the weak coupling regime with small detuning and weak field amplitude. For strong- and ultrastrong-coupling physics \cite{Liu8654003,Ashhab042311,Bourassa032109}, the RWA is no longer valid and may lead to faulty results \cite{Larson033601,Sun012107}. In paticular, developments in physical implementation lead to strong coupling between qubit and cavity modes \cite{You474589,Niemczyk6772}, which requires a careful consideration of the effect of counter-rotating terms. This motivates us to study the shortcuts to adiabaticity technique beyond the RWA. We should stress here that the original Hamiltonian ${H_{0}}(t)$ presented in Eq.~(\ref{tilde_H0_model}) is beyond the RWA. However, we can easily obtain the counteradiabatic term for the RWA case by ignoring the counter-rotating terms ${e^{ \pm 2i{\omega _L}t}}$ in ${H_{0}}\left( t \right)$. Consequently, according to Eq.~(\ref{tilde_H0_model}) the relevant transitionless quantum driving algorithm is designed by using the RWA Hamiltonian
\begin{equation}
H_0^{RWA}(t) = \hbar \left( {\begin{array}{*{20}{c}}
{\frac{1}{2}[ - \Delta (t) + i{\gamma _1}(t)]}&{\frac{1}{2}{\Omega _R}\left( t \right) + J}\\
{\frac{1}{2}{\Omega _R}\left( t \right) + J}&{\frac{1}{2}[\Delta (t) - i{\gamma _2}(t)]}
\end{array}} \right),
\label{H0_model_RWA}
\end{equation}
where ${\Omega _R}(t)$ is the Rabi frequency and ${\Delta (t)}$ is the detuning from the atomic time-dependent transition frequency ${\omega _0}$. We choose a linearly chirped Gaussian pulse with detuning $\Delta (t) =  - 2{y_1}t$ and Gaussian Rabi frequency ${\Omega _R}(t) = {\Omega _0}{e^{ - {x_1}{t^2}}}$.
Now using Eqs.~(\ref{general_H0}), (\ref{H1_general}) and (\ref{H0_model_RWA}), we can obtain the corresponding exact counteradiabatic driving Hamiltonian $H_{1}^{RWA}$$(t)$  with RWA. In Fig.~\ref{population_CR_new}(a) and~\ref{population_CR_new}(b), we plot the populations ${P_1}\left( t \right)$ and ${P_2}\left( t \right)$ with RWA, where $J = 0.05i$ GHz. We show that all the counter-rotating terms are neglected in Fig.~\ref{population_CR_new}(a), but they are considered in Fig.~\ref{population_CR_new}(b). In the RWA regime, we use Hamiltonian $H_{0}^{RWA}(t)$ and the counteradiabatic driving Hamiltonian $H_{1}^{RWA}(t)$ to describe the two-level system (see Fig.~\ref{population_CR_new}(a)). We find that the highly efficient transfer of population from the excited state to the ground state can be achieved. However, the effects of the counter-rotating terms are manifestly strong and should not be neglected when the RWA fails. At this time, $H_{1}^{RWA}(t)$ can not be used as the counteradiabatic driving Hamiltonian to accelerate the adiabatic dynamics governed by the reference Hamiltonian ${H_{0}}(t)$ in Eq.~(\ref{tilde_H0_model}) (see Fig.~\ref{population_CR_new}(b)), which suggests that all the counter-rotating terms should be considered in the present construction of the desired transitionless drivings. Thus, in the regime beyond the RWA, we have to redesign the counteradiabatic terms of the shortcuts to adiabaticity according to Eq.~(\ref{H1_general}), which can be given by Eq.~(\ref{tildeH1}). The time evolutions of the populations ${P_1}\left( t \right)$ and ${P_2}\left( t \right)$ under the control of the total Hamiltonian ${H}(t)={H_{0}}(t)+{H_{1}}(t)$ with ${H_{0}}(t)$ and ${H_{1}}(t)$ respectively given by Eqs.~(\ref{tilde_H0_model}) and (\ref{tildeH1}) are shown in Fig.~\ref{population_CR_new}(c), which describes the case beyond the RWA. From Fig.~\ref{fidelity_CR_new}(b) and~\ref{fidelity_CR_new}(c), we see that the fidelity of transitionless population transfers implemented by the shortcuts to adiabaticity technique beyond the RWA can be enhanced remarkably, which indicates again that the previous shortcuts to adiabaticity technique (under the RWA) does not work when RWA fails. It is seen further from Fig.~\ref{fidelity_CR_new} that the fidelity is higher than $90\% $ if the population transfer is implemented by the shortcuts to adiabaticity technique beyond the RWA with the appropriate parameters. Different from the previous results \cite{Chen023405,Li2530135}, the shortcuts to adiabaticity technique beyond the RWA we study here is in the case of strong non-Hermiticity and reference Hamiltonian ${H_{0}}(t)$ in Eq.~(\ref{tilde_H0_model}) is a matrix with general form whose off-diagonal elements  are not conjugate to each other. This provides a wider possibility for the practical application of shortcuts to adiabaticity.

\section{Discussions on the experimental implementation}
In this section, we present discussions on the experimental implementation of the general two-level non-Hermitian system with off-diagonal elements which are not conjugate to each other. Compared to the two-level Hermitian systems, the two-level non-Hermitian systems impose a greater experimental challenge to study fast time-dependent processes that reproduce the effect of a slow, adiabatic driving of a quantum system. Since the off-diagonal terms of the Hamiltonian are not the complex conjugate to each other, in general there is no simple laser interaction leading to Hamiltonian (\ref{general_H0}). However, recently it is pointed out that these general Hamiltonians can be implemented in some systems experimentally. In the following, we provide two physical systems as examples, which can realize Eq.~(\ref{general_H0}).
%figure7
\begin{figure}[b]
\centerline{
\includegraphics[width=7.8cm, height=4.6cm, clip]{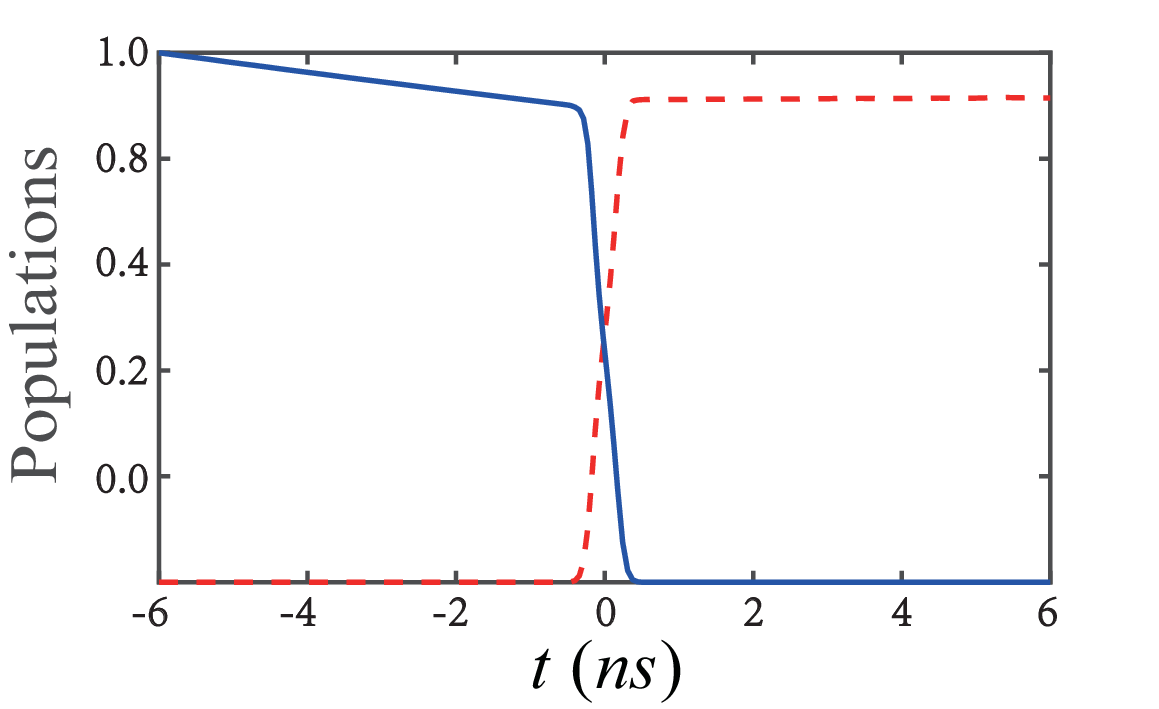}}
\caption{The evolution of the population for a two-level non-Hermitian system with gain and loss rate ${\gamma _1} = 2\pi  \times 0.1$ MHz and ${\gamma _2} = 2\pi  \times 3$ MHz, where the total Hamiltonian is ${H}(t)={H_{0}}(t)+{H_{1}}(t)$ with ${H_{0}}(t)$ and ${H_{1}}(t)$ respectively given by Eqs.~(\ref{tilde_H0_model_new}) and (\ref{H1_general}).  The red dashed and blue solid lines correspond to ${P_1}(t)$ and ${P_2}(t)$, respectively. Here we choose Gaussian pulse with detuning $\Delta (t) =  - 2{y_2}t$ and Gaussian Rabi frequency ${\Omega _R}(t) = {\Omega _0}{e^{ - {x_2}{t^2}}}$. The parameters chosen are ${x_2} = {(2\pi )^2} \times 0.3$ GHz$^2$, ${y_2} = {(2\pi )^2} \times 0.005$ GHz$^2$, and ${\Omega _0} = 2\pi  \times 0.01$ GHz. The population transfer is implemented by the shortcuts to adiabaticity technique.}\label{experiment_new}
\end{figure}

\begin{figure}[htbp]
\centerline{
\includegraphics[width=7cm, height=4.8cm, clip]{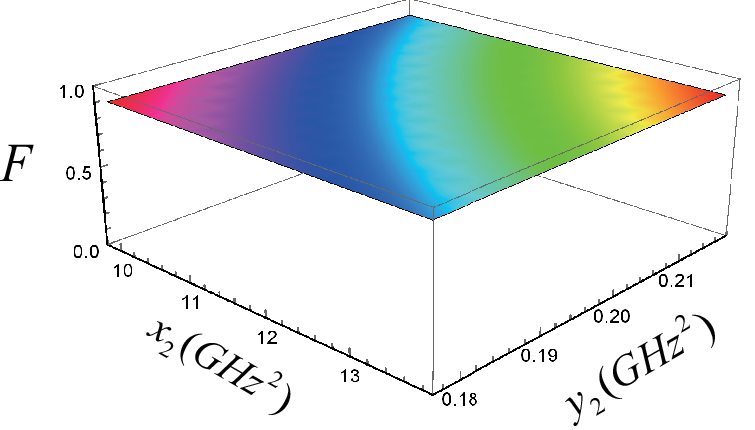}}
\caption{Parameter-dependent fidelity as a function of pulse-parameter ${x_2}$ and ${y_2}$  with ${t_f}=6$ ns. The parameters are chosen as ${x_2} = {(2\pi )^2} \times [0.25,0.35]$ GHz$^2$ and ${y_2} = {(2\pi )^2} \times [0.0045,0.0055]$ GHz$^2$. Other parameters are the same as described in the caption of Fig.~\ref{experiment_new}.}\label{fidelity_experiment_new2}
\end{figure}

\emph{Atom-light interaction systems}. Applying the electric dipole approximation and a laser-adapted interaction picture, the general Hamiltonian with disregarding atomic motion reads
\begin{eqnarray}
\begin{aligned}
{H_{0}}(t) = \frac{\hbar }{2}\left( {\begin{array}{*{20}{c}}
{ - \Delta (t) + i{\gamma _1}(t)}&{\Omega (t){e^{i{\theta _1}}}}\\
{{\Omega ^*}(t){e^{i{\theta _2}}}}&{\Delta (t) - i{\gamma _2}(t)}
\end{array}} \right),
\label{tilde_H0_experiment}
\end{aligned}
\end{eqnarray}
where $\Omega (t) = {\Omega _R}(t){e^{ - i{\omega _L}t}}$ and $\Delta (t) = {\omega _0}(t) - {\omega _L}$. By selecting the appropriate parameters, e.g., ${\theta _1} = {\theta _2} = 2 + \frac{1}{2}\pi $ and using Gaussian pulse with Rabi frequency ${\Omega _R}(t) = {\Omega _0}{e^{ - {x_2}{t^2}}}$ and detuning $\Delta (t) =  - 2{y_2}t$, we can get the general Hamiltonian that we desired whose off-diagonal elements are not conjugate to each other, which can be realized in the driven system of the atom-light interaction \cite{Peng123842,Lai57665,Holler1809,Wang111610}. From Ref.~\cite{Peng123842}, we can see that the off-diagonal elements of the effective Hamiltonian which is obtained to govern the dynamics of the two collective spin-wave excitations are non-Hermitian under certain approximations. Substituting these two chosen parameters ${\theta _1}$ and ${\theta _2}$ into Eq.~(\ref{tilde_H0_experiment}), we have \cite{Peng123842,Lai57665,Holler1809,Wang111610}
\begin{equation}
\begin{aligned}
{H_{0}}(t) = \frac{\hbar }{2}\left( {\begin{array}{*{20}{c}}
{ - \Delta (t) + i{\gamma _1}(t)}&{i{\Omega _R}(t){e^{ - i{\omega _L}t}}{e^{2i}}}\\
{i{\Omega _R}(t){e^{i{\omega _L}t}}{e^{2i}}}&{\Delta (t) - i{\gamma _2}(t)}
\end{array}} \right),
\label{tilde_H0_model_new}
\end{aligned}
\end{equation}\\
where ${\Omega _R}(t)$  is the Rabi frequency and ${\Delta (t)}$ is the detuning from the atomic time-dependent transition frequency ${\omega _0}$. According to Eq.~(\ref{H1_general}), the corresponding counteradiabatic driving Hamiltonian ${{H}_{1}}(t)$ can also be obtained. The time evolution of the populations governed by the total Hamiltonian ${H}\left( t \right) = {{H}_{0}}\left( t \right) + {{H}_{1}}\left( t \right)$ is shown in Fig.~\ref{experiment_new}. We can see that under the control of total Hamiltonian ${{H}}(t)$, the intended fast population inversion can be realized successfully. In Fig.~\ref{fidelity_experiment_new2}, we show the fidelity in order to characterize the transitonless population transfer efficiency. We can clearly see that the fidelity is insensitive to the fluctuations of both parameters ${x_2}$ and ${y_2}$, which indicates fast population inversion for a wide range of parameters could be obtained and the shortcuts to adiabaticity could be constructed availability.

\emph{Whispering-gallery microcavity systems}. The general Hamiltonian with off-diagonal elements that are not conjugate to each other can also be obtained in the two-mode-approximation model which is based on a two-mode approximation for counter-travelling waves in a whispering-gallery microcavity \cite{Chen548192}, such as a micro-disk or micro-toroid, perturbed by $N$ Rayleigh scatterers. The effective $2 \times 2$ Hamiltonian in the travelling-wave basis (counterclockwise (CCW), clockwise (CW)) is
\begin{equation}
\begin{aligned}
{H^{\left( M \right)}}{\rm{ = }}\left( {\begin{array}{*{20}{c}}
{{\Omega ^{\left( M \right)}}}&{{A^{\left( M \right)}}}\\
{{B^{\left( M \right)}}}&{{\Omega ^{\left( M \right)}}}
\end{array}} \right)
\label{HN}
\end{aligned}
\end{equation}
where ${\Omega ^{( M )}} = {\Omega ^{(0)}} + \sum\nolimits_{j = 1}^M {({V_j} + {U_j})} $, ${A^{( M )}} = \sum\nolimits_{j = 1}^M {({V_j} - {U_j}){e^{ - i2m{\beta _j}}}} $ and ${B^{( M )}} = \sum\nolimits_{j = 1}^M {({V_j} - {U_j}){e^{i2m{\beta _j}}}}$ with $m$ is the azimuthal mode number. ${\Omega ^{\left( 0 \right)}}$ is the complex frequency of the unperturbed resonance mode. ${\beta _j}$ is the azimuthal position of the $j$th nanoparticle. The complex numbers $2{V_j}$ and $2{U_j}$ are frequency shifts for positive- and negative-parity modes introduced by local perturbation $j$ alone. The Hamiltonian in Eq.~(\ref{HN}) is general non-Hermitian because ${\Omega ^{\left( 0 \right)}}$, ${{V_j}}$ and ${{U_j}}$ are complex numbers. While the diagonal elements describe the mean frequency and decay rate, the off-diagonal elements describe the coherent backscattering of light from CCW to CW (${B^{\left( N \right)}}$) and from CW to CCW (${A^{\left( N \right)}}$) propagation direction. In particular, it is general that $\left| {{A^{\left( N \right)}}} \right| \ne \left| {{B^{\left( N \right)}}} \right|$, that is, the backscattering between CW and CCW traveling waves is asymmetric \cite{Wiersig063828,Wiersig012119,Wiersig203901}. It has been demonstrated that the two-state model (\ref{HN}) works very well for a disk with a few (but not too many) external scatterers \cite{Wiersig063828}. According to the above analysis, we can find that the target Hamiltonian whose off-diagonal elements are not conjugate to each other may be implemented experimentally.

%figure8

\section{Conclusion}
In conclusion, we have generalized the shortcuts to adiabaticity technique for general non-Hermitian Hamiltonian systems and provided application examples. We numerically and analytically study the influence caused by perturbation $J$ of pulse and the effects caused by off-diagonal elements which are not conjugate to each other in both weak non-hermiticity and strong non-hermiticity regimes, respectively. Different from the previous results, we found that the shortcuts to adiabaticity in the strong non-hermiticity regime can also be realized theoretically and experimentally. Meanwhile, using the general two-level non-Hermitian Hamiltonian, we have applied this method to the linearly chirped Gaussian pulse model as an application example, where the results of numerical simulation show that the previous shortcuts to adiabaticity technique with RWA does not work well if the counter-rotating terms can not be neglected. Compared with the adiabatic protocol, our modified method of shortcuts to adiabaticity is faster as it can drive a system from a given initial state to a prescribed final state, which reduces more decoherence in the whole evolution process of the quantum system. Moreover, with shorter process times, the experiments can be repeated more often to increase signal-to-noise ratios. In particular, it might play an important role, e.g., in performing a quantum algorithm by a series of continuous quantum coherent manipulations with a driven quantum system whose coherence lifetime is limited.

Quantum dynamics of systems governed by non-Hermitian Hamiltonians is currently a very popular field of research. An open problem is to extend the present general model to the study of non-Hermitian systems coupled with a dissipative environment, where a normalized Lindblad master equation needs to be re-derived \cite{Brody230405,Sergi062108,Zloshchastiev611298}. In addition, assessing the cost of implementing shortcuts to adiabaticity arises as a natural question with both fundamental and practical implications in nonequilibrium statistical mechanics, which has already been discussed for unitary systems \cite{Campbell100601,Funo100602,Torrontegui022133}. Therefore, exploring the time-energy cost and quantum speed-limit for our non-Hermitian shortcuts to adiabaticity is also an interesting subject in the future work.

\section*{ACKNOWLEDGMENTS}

This work is supported by National Natural Science Foundation of China (NSFC) under Grants No.~11775048 and No.~12047566, Fundamental Research Funds for the Central Universities under Grant No.~2412019FZ044, and Natural Science Foundation of Jilin Province (subject arrangement project) under Grant No. 20210101406JC.

\section*{}
\appendix
\section{\label{APPA} The instantaneous eigenvalues and corresponding instantaneous eigenstates of Eq.~(\ref{general_H0})}
The dynamics of the general two-level non-Hermitian system determined by Eq.~(\ref{general_H0}) is described exactly by the Schr${\rm{\ddot o}}$dinger equation
\begin{eqnarray}
\begin{aligned}
i\hbar | {\dot \Psi (t)} \rangle  = {H_0}\left( t \right)\left| {\Psi (t)} \right\rangle,
\label{SchH0}
\end{aligned}
\end{eqnarray}
with $\left| {\Psi (t)} \right\rangle =(\begin{array}{*{20}{c}}
a(t)&b(t)\end{array})^T$, where $a\left( t \right)$ and $b\left( t \right)$ denote probability amplitudes of the two bare states $\left| 1 \right\rangle  = {(\begin{array}{*{20}{c}}
1&0\end{array})^T}$ and $\left| 2 \right\rangle  = {(\begin{array}{*{20}{c}}
0&1\end{array})^T}$, respectively. $T$ denotes transposition and dot denotes the derivative with respect to time. According to the eigenequation ${H_0}(t)\left| {{E_n}(t)} \right\rangle  = {E_n}(t)\left| {{E_n}(t)} \right\rangle $, we can get the instantaneous eigenvalues for Hamiltonian (\ref{general_H0}) as follows
\begin{eqnarray}
\begin{aligned}
{E_ \pm }\left( t \right) = \hbar[k\left( t \right) \pm \sqrt {{d^2}\left( t \right) + g\left( t \right)\Omega \left( t \right)}]
\label{Edg},
\end{aligned}
\end{eqnarray}
where $d\left( t \right) = \frac{1}{2}\left[ {e\left( t \right) - h\left( t \right)} \right]$, $k\left( t \right) = \frac{1}{2}\left[ {e\left( t \right) + h\left( t \right)} \right]$, and corresponding eigenstates are
\begin{eqnarray}
\left| {{E _ + }\left( t \right)} \right\rangle  &=& \frac{1}{{{S_ + }\left( t \right)}}\left( {\begin{array}{*{20}{c}}
{\Omega \left( t \right)}\\
{{\Lambda _ + }\left( t \right)}
\end{array}} \right),\\
\left| {{E _ - }\left( t \right)} \right\rangle  &=& \frac{1}{{{S_ - }\left( t \right)}}\left( {\begin{array}{*{20}{c}}
{\Omega \left( t \right)}\\
{{\Lambda _ - }\left( t \right)}
\end{array}} \right),
\label{Psi_f}
\end{eqnarray}
with
\begin{equation}
\begin{aligned}
{\Lambda _ \pm }\left( t \right) =& d\left( t \right) \pm \sqrt {{d^2}\left( t \right) + g\left( t \right)\Omega \left( t \right)},\\
{S_ \pm }\left( t \right) =& \sqrt {\Omega \left( t \right)g\left( t \right) + {\Lambda _ \pm^2 }\left( t \right)}.
\label{Edgzz}
\end{aligned}
\end{equation}
The adjoint of  ${H_0}\left( t \right)$ reads
\begin{eqnarray}
\begin{aligned}
H_0^\dag \left( t \right) =\hbar \left( {\begin{array}{*{20}{c}}
{{h^ * }\left( t \right)}&{{g^ * }\left( t \right)}\\
{{\Omega ^ * }\left( t \right)}&{{e^ * }\left( t \right)}
\end{array}} \right),
\label{H0_dagger}
\end{aligned}
\end{eqnarray}
and corresponding Schr${\rm{\ddot o}}$dinger equation satisfying $\langle {\tilde \Psi (t)}\left| {\Psi (t)} \right\rangle = 1$ is
\begin{eqnarray}
\begin{aligned}
i\hbar |\dot{\tilde \Psi} (t)\rangle  = H_0^\dag (t)|\tilde \Psi (t)\rangle,
\label{S_eq_H0_dagger}
\end{aligned}
\end{eqnarray}
where the dagger denotes Hermitian conjugation. The eigenequation for $H_0^\dag (t)$ is $H_0^\dag (t)|{{\tilde E}_n}(t)\rangle  = E_n^*(t)|{{\tilde E}_n}(t)\rangle $, where the asterisk denotes complex conjugate, and $|{{\tilde E}_n}(t)\rangle$ denotes the instantaneous eigenstate of $H_0^\dag (t)$, which satisfies the orthogonal and normalized  relations as $\langle {{{\tilde E}_n}(t)} | {{E_m}(t)} \rangle  = {\delta _{nm}}$ and $\sum\nolimits_n {| {{{\tilde E}_n}(t)} \rangle } \langle {{E_n}(t)} | = \sum\nolimits_n {| {{E_n}(t)} \rangle } \langle {{{\tilde E}_n}(t)} | = 1$. The left instantaneous eigenstates of ${H_0}\left( t \right)$ read
\begin{eqnarray}
\langle {{{\tilde E }_ + }\left( t \right)} | &=& \frac{1}{{{S_ + }\left( t \right)}}\left( {\begin{array}{*{20}{c}}
{g\left( t \right)}&{{\Lambda _ + }\left( t \right)}
\end{array}} \right),\label{Psi_hat_z}\\
\langle {{{\tilde E }_ - }\left( t \right)} | &=& \frac{1}{{{S_ - }\left( t \right)}}\left( {\begin{array}{*{20}{c}}
{g\left( t \right)}&{{\Lambda _ - }\left( t \right)}
\end{array}} \right).
\label{Psi_hat_f}
\end{eqnarray}

\section{\label{APPB} The adiabaticity condition for general non-Hermitian Hamiltonian (\ref{general_H0})}
The adiabaticity condition \cite{Ibanez023415} for Eq.~(\ref{solution_step}) in time-dependent non-Hermitian two-level Hamiltonian (\ref{general_H0}) is given by
\begin{equation}
\begin{aligned}
\frac{{\hbar \left| {\langle {{\tilde E }_ + }(t)|{{\dot E }_ - }(t)\rangle } \right|{e^{ - {\rm{Im\{ }}\frac{1}{\hbar }\int_0^t {[{E_ + }(t_1) - {E_ - }(t_1)]d{t_1}\} } }}}}{{\left| {{E_ + }(t) - {E_ - }(t)} \right|}} \ll 1
\label{general_condition},
\end{aligned}
\end{equation}
which is a general adiabatic condition without being limited to the weak non-Hermiticity regime discussed in detail in Sec. III.

Substituting Eqs.~(\ref{Edg}), (\ref{Psi_f}) and (\ref{Psi_hat_z}) into Eq.~(\ref{general_condition}), we obtain
\begin{equation}
\begin{aligned}
\frac{{\left| {s(t)} \right|{e^{ - {\rm{Im}}[\int_0^t {2 \sqrt {{d^2}({t_1}) + g({t_1})\Omega ({t_1})} d{t_1}]} }}}}{{\left| {2 \sqrt {{d^2}(t) + g(t)\Omega (t)} } \right|}} \ll 1
\label{conditionform},
\end{aligned}
\end{equation}
where $s(t) = \frac{1}{{{S_ + }(t)}}[g(t)\Omega (t) + {\Lambda _ + }(t){\Lambda _ - }(t)]\frac{d}{{dt}}(\frac{1}{{{S_ - }(t)}}) + \frac{1}{{{S_ + }(t){S_ - }(t)}}[g(t)\dot \Omega (t) + {\Lambda _ + }(t){{\dot \Lambda }_ - }(t)]$.

\section{\label{APPC} The derivation of Eq.~(\ref{H1_perturbation_original})}
Substituting $\Omega (t) = K(t) + {J_0}(t)$ and $g(t) = {K^ * }(t) + {J_1}(t)$ into Eq.~(\ref{H1_general}), we can obtain
\begin{equation}
\begin{aligned}
A(t) =& \dot K(t)[e(t) - h(t)] - K[\dot e(t) - \dot h(t)] &\\
&+ {{\dot J}_0}(t)[e(t) - h(t)] - {J_0}(t)[\dot e(t) - \dot h(t)],&\\
D(t) =& {{\dot K}^ * }(t)[e(t) - h(t)] - {K^ * }(t)[\dot e(t) - \dot h(t)] &\\
&+ {{\dot J}_1}(t)[e(t) - h(t)] - {J_1}(t)[\dot e(t) - \dot h(t)],&\\
{B_1}(t) =& K(t){{\dot K}^ * }(t) - \dot K(t){K^ * }(t) + K(t){{\dot J}_1}(t) &\\
&+ {{\dot K}^ *(t)}{J_0}(t) - \dot K(t){{\dot J}_0}(t) - \dot K(t){J_1}(t) &\\
&+ {J_0}(t){{\dot J}_1}(t) - {{\dot J}_0}(t){J_1}(t),&\\
{C_1}(t) =& {d^2}(t) + {\left| K(t) \right|^2} + K(t){J_1}(t) + {K^ * }(t){J_0}(t) &\\
&+ {J_0}(t){J_1}(t).
\label{matrix}
\end{aligned}
\end{equation}
Therefore, based on the series expansion, we can obtain
\begin{equation}
\begin{aligned}
\frac{1}{{{C_1}(t)}} =& \frac{1}{{{{\left| K(t) \right|}^2} + {d^2}(t)}} \cdot \frac{1}{{1 + z(t)}}\\
\approx &\frac{1}{{{{\left| K(t) \right|}^2} + {d^2}(t)}}[1 - z(t)]\\
=& \frac{1}{{{{\left| K(t) \right|}^2} + {d^2}(t)}} - \frac{{K(t){J_1}(t)}}{{{{[{{\left| K(t) \right|}^2} + {d^2}(t)]}^2}}}\\
&+ \frac{{{K^ * }(t){J_0}(t)}}{{{{[{{\left| K(t) \right|}^2} + {d^2}(t)]}^2}}} + \frac{{{J_0}(t){J_1}(t)}}{{{{[{{\left| K(t) \right|}^2} + {d^2}(t)]}^2}}},
\label{C12}
\end{aligned}
\end{equation}
when $\left| z(t) \right| \ll 1$, where we have set $$z(t) = \frac{{K(t){J_1}(t)}}{{{{\left| K(t) \right|}^2} + {d^2}(t)}} + \frac{{{K^ * }(t){J_0}(t)}}{{{{\left| K(t) \right|}^2} + {d^2}(t)}} + \frac{{{J_0}(t){J_1}(t)}}{{{{\left| K(t) \right|}^2} + {d^2}(t)}}.$$ Substituting Eq.~(\ref{C12}) into Eq.~(\ref{H1_general}), we can obtain the approximate counteradiabatic driving Hamiltonian (\ref{H1_perturbation_original}), where Eq.~(\ref{C12}) ignores the second-order and higher-order terms of the parameter $z(t)$.

\section{\label{APPD} The derivation of Eq.~(\ref{H1expand1})}
In order to further discuss the difference and relationship between our results and previous ones, we assume
\begin{equation}
\begin{aligned}
\Omega \left( t \right) = & K\left( t \right){\rm{ + }}J,\\
g\left( t \right) =& {K^ * }\left( t \right) + J,
\label{Omegag}
\end{aligned}
\end{equation}
where $K\left( t \right)$ is a time-dependent complex function, and $J$ is a time-independent complex number. According to Eq.~(\ref{H1_general}), we can get
\begin{small}
\begin{equation}
\begin{aligned}
\frac{1}{{{C_1}(t)}} =& \frac{1}{{{J^2} + 2{K_r}(t)J + {{\left| {K(t)} \right|}^2} + {d^2}(t)}}\\
 =& \frac{1}{{{s_1}(t) - {s_2}(t)}} \cdot \frac{1}{{J - {s_1}(t)}} + \frac{1}{{{s_2}(t) - {s_1}(t)}} \cdot \frac{1}{{J - {s_2}(t)}}\\
 =& \frac{{ - 1}}{{{s_1}(t)[{s_1}(t) - {s_2}(t)]}} \cdot \sum\limits_{n = 0}^\infty  {{{[\frac{J}{{{s_1}(t)}}]}^n}}\\
  &+ \frac{1}{{{s_2}(t)[{s_1}(t) - {s_2}(t)]}}\cdot \sum\limits_{n = 0}^\infty  {{{[\frac{J}{{{s_2}(t)}}]}^n}}\\
 =& \frac{1}{{{s_1}(t) - {s_2}(t)}}\sum\limits_{n = 0}^\infty  {{J^n}[\frac{1}{{s_2^{n + 1}(t)}} - \frac{1}{{s_1^{n + 1}(t)}}]},
\label{C1}
\end{aligned}
\end{equation}
\end{small}
which holds valid for $| J | < \min \{ | {{s_1}(t)} |,| {{s_2}(t)} |\} $ with ${s_1}(t) =  - {K_r}(t) + \sqrt {K_r^2(t) - [{{| {K(t)} |}^2} + {d^2}(t)]} $ and ${s_2}(t) =  - {K_r}(t) - \sqrt {K_r^2(t) - [{{| {K(t)} |}^2} + {d^2}(t)]}$ being the roots of ${C_1}(t)=0$, where ${K_r}(t) = \frac{1}{2}[K(t) + {K^ * }(t)]$ denotes the real part of $K(t)$.
Substituting Eqs.~(\ref{Omegag}) and (\ref{C1}) into Eq.~(\ref{H1_general}), we can obtain the exact counteradiabatic driving Hamiltonian
\begin{widetext}
\begin{equation}
\begin{aligned}
{H_1}(t)=& \frac{{i\hbar }}{{4[{s_1}(t) - {s_2}(t)]}}\sum\limits_{n = 0}^\infty  {{J^n}} [\frac{1}{{s_2^{n + 1}(t)}} - \frac{1}{{s_1^{n + 1}(t)}}]&\\
&\times \left( {\begin{array}{*{20}{c}}
{\dot K(t){K^ * }(t) - K(t){{\dot K}^ * }(t)}&{\dot K(t)[e(t) - h(t)] - K(t)[\dot e(t) - \dot h(t)]}\\
{{K^ * }(t)[\dot e(t) - \dot h(t)] - {{\dot K}^ * }(t)[e(t) - h(t)]}&{K(t){{\dot K}^ * }(t) - \dot K(t){K^ * }(t)}
\end{array}} \right)&\\
&+ \frac{{i\hbar }}{{4[{s_1}(t) - {s_2}(t)]}}\sum\limits_{n = 0}^\infty  {{J^{n + 1}}} [\frac{1}{{s_2^{n + 1}(t)}} - \frac{1}{{s_1^{n + 1}(t)}}]\left( {\begin{array}{*{20}{c}}
{ - {{\dot K}^ * }(t) + \dot K(t)}&{ - [\dot e(t) - \dot h(t)]}\\
{\dot e(t) - \dot h(t)}&{{{\dot K}^ * }(t) - \dot K(t)}
\end{array}} \right),&
\label{H1_generalapp}
\end{aligned}
\end{equation}
\end{widetext}
which can be rewritten as the form of Eq.~(\ref{H1expand1}), where
\begin{widetext}
\begin{small}
\begin{equation}
\begin{aligned}
L^{(n)}(t) =& \frac{{i\hbar }}{{4[{s_1}(t) - {s_2}(t)]}}[\frac{1}{{s_2^{n + 1}(t)}} - \frac{1}{{s_1^{n + 1}(t)}}]\left( {\begin{array}{*{20}{c}}
{\dot K(t){K^*}(t) - K(t){{\dot K}^*}(t)}&{\dot K(t)[e(t) - h(t)] - K(t)[\dot e(t) - \dot h(t)]}\\
{{K^*}(t)[\dot e(t) - \dot h(t)] - {{\dot K}^*}(t)[e(t) - h(t)]}&{K(t){{\dot K}^*}(t) - \dot K(t){K^*}(t)}
\end{array}} \right)&\\
& + \frac{{i\hbar }}{{4[{s_1}(t) - {s_2}(t)]}}[\frac{1}{{s_2^n(t)}} - \frac{1}{{s_1^n(t)}}]\left( {\begin{array}{*{20}{c}}
{ - {{\dot K}^*}(t) + \dot K(t)}&{ - [\dot e(t) - \dot h(t)]}\\
{\dot e(t) - \dot h(t)}&{{{\dot K}^*}(t) - \dot K(t)}
\end{array}} \right).&\label{H1expandN}
\end{aligned}
\end{equation}
\end{small}
\end{widetext}
Finally, we can obtain Eq.~(\ref{H1011z}) according to Eqs.~(\ref{cN}) and (\ref{H1lim}) with $N=1$, where the zeroth- and first-order expansions of ${H_{1}^\prime}(t)$ given by Eq.~(\ref{H1011z}) are as follows:
\begin{widetext}
\begin{equation}
\begin{aligned}
L^{(0)}(t) =& \frac{{i\hbar }}{{4{s_1}(t){s_2}(t)}}\left( {\begin{array}{*{20}{c}}
{\dot K(t){K^ * }(t) - K(t){{\dot K}^ * }(t)}&{\dot K(t)[e(t) - h(t)] - K(t)[\dot e(t) - \dot h(t)]}\\
{{K^ * }(t)[\dot e(t) - \dot h(t)] - {{\dot K}^ * }(t)[e(t) - h(t)]}&{K(t){{\dot K}^ * }(t) - \dot K(t){K^ * }(t)}
\end{array}} \right),\\
L^{(1)}(t) =& \frac{{i\hbar }}{{4{s_1}(t){s_2}(t)}}\left( {\begin{array}{*{20}{c}}
{ - {{\dot K}^ * }(t) + \dot K(t)}&{ - [\dot e(t) - \dot h(t)]}\\
{\dot e(t) - \dot h(t)}&{{{\dot K}^ * }(t) - \dot K(t)}
\end{array}} \right)&\\
& + \frac{{i\hbar [{s_1}(t) + {s_2}(t)]}}{{4s_1^2(t)s_2^2(t)}}\left( {\begin{array}{*{20}{c}}
{\dot K(t){K^ * }(t) - K(t){{\dot K}^ * }(t)}&{\dot K(t)[e(t) - h(t)] - K(t)[\dot e(t) - \dot h(t)]}\\
{{K^ * }(t)[\dot e(t) - \dot h(t)] - {{\dot K}^ * }(t)[e(t) - h(t)]}&{K(t){{\dot K}^ * }(t) - \dot K(t){K^ * }(t)}
\end{array}} \right).
\label{H1011}
\end{aligned}
\end{equation}
\end{widetext}

\end{document}